\documentclass[10pt,twoside]{article}
\usepackage{fancyhdr}
\usepackage[colorlinks,citecolor=blue,urlcolor=blue,linkcolor=blue,bookmarks=false]{hyperref}
\usepackage{amsfonts,epsfig,graphicx}
\usepackage{afterpage}
\usepackage{amsmath,amssymb,amsthm,dsfont}
\usepackage{fullpage}
\usepackage[T1]{fontenc} 
\usepackage{epsf} 
\usepackage{graphics}
\usepackage{xcolor}
\usepackage{comment}
\usepackage[mode=buildnew]{standalone}
\usepackage{amsfonts,amsmath}
\usepackage[sort]{natbib} 
\usepackage{psfrag,xspace}
\usepackage{color,etoolbox}
\usepackage{authblk}
\usepackage{tikz}
\usetikzlibrary{backgrounds,positioning, shapes}

\def\ind{\perp\!\!\!\perp}
\def\T{{ \mathrm{\scriptscriptstyle T} }}

\newcommand{\var}{\text{var}}

\newcommand{\Qb}{\mathbb{Q}}
\newcommand{\Pb}{\mathbb{P}}
\newcommand{\Pn}{\mathbb{P}_n}
\newcommand{\Qn}{\mathbb{Q}_N}

\newcommand{\E}{\mathbb{E}}
\newcommand{\R}{\mathbb{R}}
\newcommand{\Em}{\mathcal{E}}

\newcommand{\bY}{\mathbf{Y}}
\newcommand{\by}{\mathbf{y}}
\newcommand{\bX}{\mathbf{X}}
\newcommand{\bZ}{\mathbf{Z}}
\newcommand{\bx}{\mathbf{x}}
\newcommand{\bz}{\mathbf{z}}
\newcommand{\bH}{\mathbf{H}}

\newcommand{\bzero}{\mathbf{0}}

\def\logit{\text{logit}}

\def\expit{\text{expit}}

\DeclareSymbolFont{bbold}{U}{bbold}{m}{n}
\DeclareSymbolFontAlphabet{\mathbbold}{bbold}
\newcommand{\one}{\mathbbold{1}}
\newtheorem{theorem}{Theorem}
\newtheorem{lemma}{Lemma}
\newtheorem{corollary}{Corollary}
\newtheorem{algorithm}{Algorithm}
\newtheorem{proposition}{Proposition}
\theoremstyle{definition}

\theoremstyle{remark}
\newtheorem{assumption}{Assumption}

\newtheorem{remark}{Remark}

\begin{document}
	\allowdisplaybreaks[0]
	\def\spacingset#1{\renewcommand{\baselinestretch}%
		{#1}\small\normalsize} \spacingset{1}
	
	\title{ Doubly robust capture-recapture methods \\
		for estimating population size }
	\author[1]{Manjari Das}
	\author[1]{Edward H. Kennedy}
	\author[2,3]{Nicholas P. Jewell}
	\affil[1]{Department of Statistics \& Data Science,
		Carnegie Mellon University, Pittsburgh PA}
	\affil[2]{Department of Medical Statistics, London School of Hygiene \& Tropical Medicine, London, United Kingdom}
	\affil[3]{Division of Epidemiology \& Biostatistics, School of Public Health, University of California, Berkeley CA}
	\date{}
	\maketitle
	\thispagestyle{empty}
	
	\begin{abstract}
		Estimation of population size using incomplete lists (also called the capture-recapture problem) has a long history across many biological and social sciences. For example, human rights groups often construct partial and overlapping lists of victims of armed conflicts, with the hope of using this information to estimate the total number of victims. Earlier statistical methods for this setup either use potentially restrictive parametric assumptions, or else rely on typically suboptimal plug-in-type nonparametric estimators; however, both approaches can lead to substantial bias, the former via model misspecification and the latter via smoothing. Under an identifying assumption that two lists are conditionally independent given measured covariate information, we make several contributions. First we derive the nonparametric efficiency bound for estimating the capture probability, which indicates the best possible performance of any estimator, and sheds light on the statistical limits of capture-recapture methods. Then we present a new estimator, and study its finite-sample properties, showing that it has a double robustness property new to capture-recapture, and that it is near-optimal in a non-asymptotic sense, under relatively mild nonparametric conditions. Next, we give a method for constructing confidence intervals for total population size from generic capture probability estimators, and prove non-asymptotic near-validity. Finally, we study our methods in simulations, and apply them to estimate the number of killings and disappearances attributable to different groups in Peru during its internal armed conflict between 1980 and 2000.
	\end{abstract}
	
	\newpage
	\section{Introduction}
	\label{sec:intro}
	
	Capture-recapture is a study design for estimating population size when only a fraction of the population is observed. This setup arises frequently, for example in studying ecological abundance, disease prevalence, and casualties in armed conflicts. Capture-recapture has a long history, dating back to at least Graunt in the 1600s \citep{hald2003history}, who used it to estimate plague prevalence in England. Similarly, in 1802 Laplace estimated the total population of France \citep{goudie2007captures}, and \citet{petersen1896yearly}  the abundance of plaice fish. More recently, it has been used in diverse settings ranging from estimating the number of pages on the web \citep{fienberg1999classical} to the total number of victims in a war \citep{ball2003many}, among many others. \\
	
	The simplest capture-recapture setup, credited to \citet{petersen1896yearly}, consists of two independent lists with partial captures from the population of interest. There have been many generalizations over time. For our purposes, much of the previous work in capture-recapture can be viewed as falling within one of three streams. The first and oldest stream includes relatively simple data structures, e.g., involving no covariate information and relatively few lists  \citep{petersen1896yearly, schnabel1938estimation, Darroch1958indeplists}. More recent advances in this stream include \citet{burnham1979robust} and \citet{lee1994estimating}. A second stream emerged to handle more intricate data structures, e.g., complex covariate information to help account for heterogeneity/dependence, largely using model-based approaches 
	\citep{link2003nonidentifiability, carothers1973effects, Fienberg1972loglinear, tilling1999capture, pollock2002use, huggins1989statistical, alho1990logistic, yip2001unified}.
	However, the advantages of this second stream typically come at the expense of potentially restrictive parametric modeling assumptions, which when violated would induce bias. A third more recent stream addresses similar data structures as the second, but using more flexible nonparametric tools, e.g., local kernel or nonparametric Bayes or spline methods \citep{huggins2007non, huggins2011review, chen2000nonparametric, manrique2016bayesian, kurtz_2018, zwane2005population, stoklosa2012robust, yee2015vgam}. However, the work in this third stream has so far relied on interpretable but typically suboptimal plug-in estimators, which can suffer from nonparametric smoothing bias and slow rates of convergence \citep{van2003unified, van2014higher,robins2008higher}. We refer to \citet{kurtz_2018} for a more detailed review of this third stream. \\
	
	Our work takes the nonparametric perspective, but uses advances in efficiency theory to characterize optimality and improve simple plug-in estimators \citep{bickel1993efficient, van2003unified, kennedy2016semiparametric}. Under an identifying assumption that two lists are conditionally independent given measured covariate information (described in Section \ref{sec:setup_notn}), we make several contributions. 
	\begin{itemize}
		\item In Section \ref{sec:efficiency} we derive the nonparametric efficiency bound for estimating the capture probability, which indicates the best possible performance of any estimator, and sheds light on the statistical limits of capture-recapture methods. 
		\item In Section \ref{sec:estim} we present a new doubly robust estimator, and study its finite-sample error, showing it is near-optimal in a non-asymptotic sense, under mild nonparametric conditions. 
		\item In Section \ref{sec:population_size} we give a general method for constructing confidence intervals for total population size from generic capture probability estimators, and prove non-asymptotic near-validity. 
		\item In Section \ref{sec:results} we study our methods in simulations, and apply them to estimate the number of killings and disappearances attributable to different groups in Peru during its internal armed conflict between 1980 and 2000.
	\end{itemize}
	
	\pagebreak
	
	\section{Preliminaries}\label{sec:setup_notn}
	
	\subsection{Setup}
	
	Consider a finite population of $n$ individuals, where the size $n$ is unknown and to be estimated. We suppose there are $K$ different lists of individuals from this population, yielding indicators $Y_{ik} \in \{0,1\}$ of whether individual $i \in \{1,...,n\}$ appeared on list $k \in \{1,...,K\}$. We let $\bY_i=(Y_{i1},...,Y_{iK})^\T$ denote the vector indicating list membership (i.e., capture profile) information for individual $i$. For example, in the $K=2$ case, a profile $\bY_i=(1,0)^T$ would mean that individual $i$ appears on list 1 but not list 2. We consider the case where covariates $\bX_i \in \R^d$ are also available for each individual $i=1,...,n$. We assume an individual's chances of appearing on any given lists (and their covariates) do not depend on what happens with any other individuals, and also that the covariate and (conditional) list membership distributions are the same across individuals $i=1,...,n$. This implies that the random vectors $\bZ_i=(\bX_i, \bY_i)$ are independent and identically distributed according to some distribution $\Pb$. \\
	
	\begin{remark}
		The setup above is commonly referred to as ``heterogeneous'' \citep{huggins1989statistical, tilling1999capture, pollock2002use} since list membership $\bY$ can vary with covariates $\bX$. In other words, individuals with different covariates can have different chances of list membership. \\
	\end{remark}
	
	\begin{remark}
		In what follows we use the following standard notation. We let $\E_\Qb$ denote an expectation under distribution $\Qb$, and let $\| f \|_\Qb^2 = \int f(z)^2 \ d\Qb(x)$ denote the corresponding squared $L_2(\Qb)$ norm; we let $\Qn$ denote the empirical measure under distribution $\Qb$. Finally we let $a \lesssim b$ mean $a \leq Cb$ for some universal constant $C$.  \\
	\end{remark}
	
\begin{figure}[h]
		\begin{center}
			\fbox{
				\begin{tikzpicture}[background rectangle/.style={fill= lightgray!20!}, show background rectangle]
					\node[ellipse,draw, minimum width=4.5cm, minimum height=4.5cm, fill=darkgray!42!, fill opacity=.5] (1) {};
					\node[ellipse,draw, minimum width=4cm, minimum height=3.75cm, fill=darkgray!42!, fill opacity=.5] (2) [right= 1,xshift=-.1cm, yshift=-.1cm] {};
					\node[ellipse,draw, minimum width=5cm, minimum height=3cm, fill=darkgray!42!, fill opacity=.5]   (3) [below= 1,xshift=1.5cm, yshift=.6cm] {};
					\node[] (4) [left=1, xshift=-2cm, yshift=-.5cm] {Unobserved};
					\node at (4) [yshift=-.5cm] {($\bY=\bzero$)};
					\node at (1) [yshift=.25cm] {List 1};
					\node at (1) [yshift=-.25cm] {($Y_1=1$)};
					\node at (2) [yshift=.25cm] {List 2};
					\node at (2) [yshift=-.25cm] {($Y_2=1$)};
					\node at (3) [yshift=.25cm] {List 3};
					\node at (3) [yshift=-.25cm] {($Y_3=1$)};
			\end{tikzpicture}}
			\caption{Schematic of data structure for $K=3$ lists. Observed data (i.e., those with $\bY \neq \bzero$ in the union of the three lists) are represented with dark gray, while unobserved individuals with $\bY=\bzero$ are represented with light gray. Individuals appearing in all three lists have $\bY=(1,1,1)$.}
			\label{fig:venndiagram}
		\end{center}
	\end{figure}
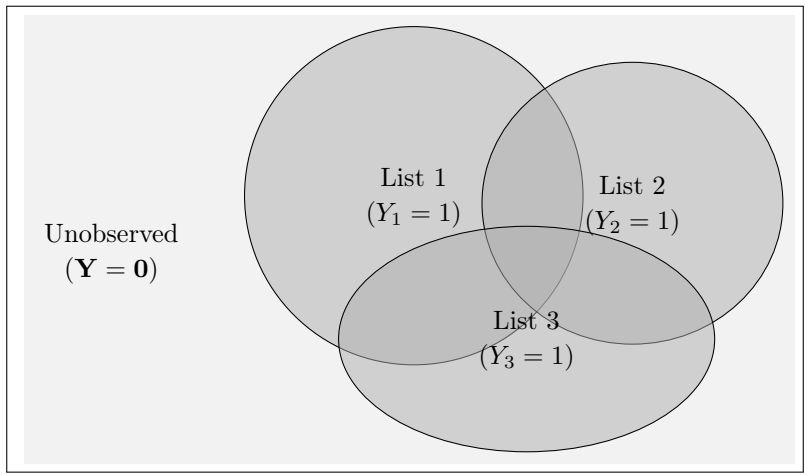

	If every individual in the population appeared on at least one list (and could be uniquely identified), then the population size would of course be known without error; however in practice a possibly substantial fraction of individuals do not appear on any list. In other words, there are some individuals with $\bY=\bzero$ that we do not observe. This means that, although the distribution $\Pb$ governs the capture profiles, we cannot sample from $\Pb$ directly. Instead we only see the $N=\sum_{i = 1}^n \one(\bY_i \neq \bzero)$ individuals for whom $Y_{ik}=1$ for some $k$. This is illustrated in Figure \ref{fig:venndiagram}.  \\
	
	\bigskip
	
	\pagebreak
	
	Hence the capture-recapture design is an example of biased sampling \citep{vardi1985empirical, breslow2000semi, qin2017biased}. In particular, the observed data $\bZ_i = (\bX_i,\bY_i), i=1,...,N$ are actually iid draws from a conditional distribution $\Qb$ defined as
	\begin{align} 
		\Qb(\bY=\by, \bX=\bx) & \equiv \Pb(\bY=\by, \bX=\bx \mid \bY \neq \bzero) \nonumber \\
		&= \psi^{-1} \Pb(\bY=\by, \bX=\bx) \ \one( \by \neq \bzero) 
	\end{align}
	where 
	\begin{equation}
		\psi = \Pb(\bY \neq \bzero)
	\end{equation}
	is the (marginal) capture probability.  \\
	
	\begin{remark}
		Some authors use $N$ to denote the total population size and $n$ for the observed number of captures; in contrast, we use $N$ for the observed number and $n$ for the total size, following the convention of saving upper case for random variables. Note the observed number of captures $N$ is random since it depends on the random selection indicators, while the total population size $n$ is fixed; specifically $N \sim \text{Bin}(n,\psi)$. Nonetheless much of our analysis will be conditional on the observed sample size $N$. \\
	\end{remark}
	
	Recall our overall goal is to estimate the total population size $n=N+\sum_i \one(\bY_i = \bzero)$. Since $N \sim \text{Bin}(n,\psi)$, the population size can be viewed as a fixed population parameter given by
	\begin{equation} \label{eq:sizefrac}
		n = \E(N) / \psi.
	\end{equation}
	Intuitively, the lower the capture probability, the more the observed number $N$ must be inflated to reflect the total population size. The quantity $\E(N)$ in \eqref{eq:sizefrac} can of course be unbiasedly estimated with the observed number of captures $N$; therefore estimating population size essentially boils down to estimating the capture probability, which can then be used to inflate the observed $N$ via the estimator
	\begin{equation}
		\widehat{n} = N / \widehat\psi.
	\end{equation}
	Thus we turn towards the crucial question of how to efficiently estimate the capture probability (specifically, in the presence of high-dimensional and/or complex covariates $\bX$), before coming back to inference about $n$ in Section \ref{sec:population_size}. In the next section we discuss identification of $\psi$ from the distribution $\Qb$ from which we sample; this will require extra assumptions, since if lists are irreparably dependent we will have no information about those who are unobserved. \\
	
	\subsection{Identification}
	\label{sec:identification}
	
	As mentioned in the previous section, without additional assumptions, the observed data distribution $\Qb$ of list membership among those on at least one list is completely uninformative about the capture probability $\psi=\Pb(\bY \neq \bzero)$. A variety of assumptions have been used to identify this and related quantities in previous literature; broadly, there must be some lack of dependence across lists in order to identify and estimate the odds and thus the overall population size. The popular Petersen estimator \citep{petersen1896yearly} assumed independence between $K=2$ lists and \citet{Darroch1958indeplists} extended it to assume independence across $K>2$ lists. \citet{Fienberg1972loglinear} assumed a log-linear model for the expected number of observations with each capture profile, with one parameter necessarily set to zero (typically the highest-order interaction term across lists). \citet{you2021estimation} have presented a generalizable framework adaptable to various identification assumptions including log-linear model assumption and independence between two lists conditional on the remaining list(s) for $K>2$ list case.\\
	
	When one has access to not only list membership but also covariate information, conditional versions of these assumptions can be used \citep{sekar1949method, chao1987estimating, tilling1999capture, huggins2007non}. Importantly, this allows heterogeneous capture probabilities that vary across units, and thus relaxes identifying assumptions; this is conceptually similar to how measured confounders are exploited in observational studies for causal inference \citep{hernan2019causal}. \citet{sekar1949method} studied the conditional case in the discrete and low-dimensional setup; the continuous case has been studied using parametric models by \citet{pledger2000unified, pollock1990statistical, tilling1999capture, huggins1989statistical, chao1987estimating, alho1990logistic}. \citet{burnham1979robust, huggins2007non} used non-parametric jackknife estimator and bandwidth selection respectively. Analogous to the assumption of \citet{tilling1999capture} for the two list case, our main identifying assumption for the $K$ list case is that there is a known pair among the $K$ lists which are conditionally independent. Without loss of generality, we let the lists be ordered so that the conditional independent pair is list 1 and 2: \\
	
	\begin{assumption}\label{as:1}
		$\Pb( Y_1 = 1 \mid \bX=\bx, Y_2=1) = \Pb(Y_1 = 1 \mid \bX=\bx, Y_2=0)$, where $Y_k$ denotes the capture indicator variable for list $k$ for $k = 1,\dots, K$. \\
	\end{assumption}
	
	Assumption \ref{as:1} says that the chance of appearing on list 1 is the same regardless of list 2 membership, among those with the same measured covariate values, i.e., that the list indicators $Y_1$ and $Y_2$ are conditionally independent given $\bX$. This assumption can be viewed as an important relaxation of a more standard assumption of marginal independence, particularly when lists cover different parts of a population.  \\
	
	For example, consider a toy setup where there are two regions, equally populous. Suppose people who live in region A have a 90\% chance of appearing on list 1 and a 10\% chance of appearing on list 2, while people who live in region B have the reverse: a 10\% chance of appearing on list 1 and a 90\% chance on list 2. Thus list 1 tends to capture region A people, and list 2 tends to capture region B people. In this case, even if conditional independence holds (i.e., the chance of appearing on list 1, within each region, is the same regardless of whether you appear on list 2), the lists will \emph{not} be marginally independent. Intuitively, the reasoning behind this is straightforward. Once we know you are on list 2, we have some additional information about your chances of appearing on list 1: you are more likely to live in region B, and so less likely to be on list 1. More specifically, the chances of appearing on either list are both 50\% marginally, but the chance of appearing on list 1 given that you are on list 2 is only 18\%. \\
	
	The conditional independence in Assumption \ref{as:1} has been used relatively extensively in the capture-recapture literature; we refer to \citet{alho1993estimating, tilling1999capture, tilling2001capture, brenner1995use, pollock1990statistical, huggins1989statistical} for more details and discussions of when this assumption may hold and when it may fail.  \\
	
	It is known \citep[e.g., as in][]{tilling1999capture} that under Assumption \ref{as:1} the capture probability $\psi=\Pb(\bY \neq \bzero)$ can be identified from the biased observed data distribution $\Qb$. Specifically, let
	\begin{align*}
		q_{1}(\bx) &= \Qb(Y_1 = 1  \mid \bX=\bx) \\
		q_{2}(\bx) &= \Qb(Y_2 = 1  \mid \bX=\bx) \\
		q_{12}(\bx) &= \Qb(Y_1 = 1, Y_2 = 1 \mid \bX=\bx) 
	\end{align*}
	denote the observational probability (under $\Qb$) of appearing on list 1, 2, and both, respectively. These probabilities will be referred to as the $q$-probabilities at various points throughout. \\
	
	\begin{remark}
		Note that when there are only $K=2$ lists, it must be that $q_1(\bx)+q_2(\bx)-q_{12}(\bx)=1$ since each observed unit must appear on list 1, list 2, or both, according to the sampling distribution $\Qb$. In general when $K>2$ one only has the requirement $0\leq q_1(\bx)+q_2(\bx)-q_{12}(\bx) \leq 1$, since it is possible some individuals only appear on lists $j \geq 3$ other than 1 and 2. \\
	\end{remark}
	
	\begin{remark}
		Note that when there are more than two lists and without loss of generality the conditionally independent list pair is 1 and 2, the remaining lists are aiding in the estimation by potentially increasing the number of observed individuals. This in turn leads to variance reduction as is discussed in appendix \ref{app:how_many_lists_to_use}.
	\end{remark}
	
	For posterity we give the identification result for $\psi$ in the following proposition (with a proof given in the Appendix).  \\
	
	\begin{proposition}\label{thm:identification}
		Under Assumption \ref{as:1} and the positivity condition $\Qb\{q_{12}(\bX)>0\} = 1$, the conditional and marginal capture probabilities are identified from $\Qb$ by
		\begin{align}
			\gamma(\bx) &\equiv {\Pb(\bY \neq \bzero \mid \bX=\bx)} = \frac{q_{12}(\bx)}{q_{1}(\bx) q_{2}(\bx)}    \\
			\psi &\equiv  {\Pb(\bY \neq \bzero)} = \left\{ \int \gamma(\bx)^{-1} \ d\Qb(\bx)  \right\}^{-1}.
			\label{eq:identification}
		\end{align}
	\end{proposition}
	
	In the following sections we give three main contributions. First we derive the efficiency bound for estimating the capture probability $\psi$ under a nonparametric model that puts no parametric restrictions on the `nuisance' functions $ (q_{1},q_{2},q_{12})$; second, we construct novel estimators that attain the efficiency bound under weak nonparametric conditions (e.g., allowing the use of flexible machine learning tools); and third, we give a general method for building corresponding confidence intervals for the total population size $n$, given any asymptotically linear estimate $\widehat\psi$ of the capture probability. \\
	
	\begin{remark}
		All subsequent results apply to the statistical parameter $\psi$, which we define from here on as the harmonic mean on the right-hand-side of \eqref{eq:identification}. Under the identifying Assumption \ref{as:1} (and positivity), $\psi$ also represents the capture probability on the left-hand-side of \eqref{eq:identification}, but our statistical results do not require this link, and apply to the harmonic mean in \eqref{eq:identification} regardless. \\
	\end{remark}
	
	\begin{remark}
		When there are more than two lists, it is possible that multiple list pairs satisfy assumption \ref{as:1}. This presents multiple different opportunities to identify and estimate the capture probability, and so yields a semiparametric model with testable implications. We leave exploration of this setup for potential future work.
	\end{remark}
	
	\section{Efficiency Bound}\label{sec:efficiency}
	
	In this section, we derive the nonparametric efficiency bound for estimating the capture probability $\psi$ using an iid sample from distribution $\Qb$. This gives a crucial benchmark against which one can compare candidate estimators: once an estimator is shown to attain this bound, no further improvements can be made (at least asymptotically) without adding extra assumptions. To the best of our knowledge, the only previous efficiency bounds in the capture-recapture literature are for low-dimensional parametric models, where standard results from maximum likelihood theory apply \citep{Fienberg1972loglinear, gimenez2005efficient}. \\
	
	In order to derive the efficiency bound, we use tools from semiparametric theory \citep{bickel1993efficient}. A fundamental goal here is to characterize influence functions, and the efficient influence function in particular. The efficient influence function of a parameter acts as the derivative term in a distributional Taylor expansion of the parameter, viewed as a map on distributions; thus it can represent the change in the parameter after perturbing the distribution it takes as input. Practically, the efficient influence function for a parameter has several critical implications. First, as mentioned above, it leads to a minimax efficiency bound \citep{van2002part} and thus provides a benchmark for efficient estimation in flexible nonparametric models. Further, it can be used to construct efficient estimators that attain the bound under weak assumptions, and sheds light on the regularity conditions necessary for said efficiency, as will be shown in Section \ref{sec:estim}. More details on nonparametric efficiency theory can be found in \citet{bickel1993efficient}, \citet{van2002part}, and \citet{van2003unified}, among others; reviews can be found in \citet{tsiatis2006semiparametric} and \citet{kennedy2016semiparametric}, for example. \\
	
	Our first result gives the form of the efficient influence function for the capture probability, in an unrestricted nonparametric model. \\
	
	\begin{lemma}\label{lem:efficient influence function}
		Let $g:\R \mapsto \R$ be any function differentiable at the true capture probability $\psi$ defined in \eqref{eq:identification}. Under a nonparametric model, the efficient influence function for the parameter $g(\psi)$  is given by $f_g(\psi) \phi(\bZ; \Qb)$ where $f_g(\psi) = -g'(\psi) \psi^2$ and 
		\begin{equation}
			\begin{aligned}
				\phi(\bZ; \Qb)
				&= \frac{1}{\gamma(\bX)} \left\{ \frac{Y_1}{q_1(\bX)} + \frac{Y_2}{q_2(\bX)} - \frac{Y_1 Y_2}{q_{12}(\bX)} \right\} - \frac{1}{\psi} .
			\end{aligned}
		\end{equation}
	\end{lemma}
	
	The proof is presented in the Appendix. Note that the first term in the efficient influence function is a product of the inverse conditional capture probability $\gamma^{-1}$ with a term whose conditional expectation given $\bX$ (under $\Qb$) equals 1. The efficient influence function will be bounded for example if $q_{12}(\bx) \geq \epsilon$ for some $\epsilon>0$ and all $\bx$ (note that $q_1(\bx) \wedge q_2(\bx) \geq q_{12}(\bx)$ so $q_{12}(\bx) \geq \epsilon$ implies all the $q$-probabilities are bounded below by $\epsilon$). \\
	
	The variance of the efficient influence function acts as a nonparametric efficiency bound, in that no estimator can achieve a better mean squared error in a local minimax sense \citep{van2002semiparametric}. The following theorem and corollary give the form of this bound and formalize the minimax result, respectively. All expectations and variances are under distribution $\Qb$ unless noted otherwise. \\
	
	\begin{theorem}\label{thm:var_phi}
		Let $g:\R \mapsto \R$ be any function differentiable at $\psi$. The nonparametric efficiency bound for estimation of $g(\psi)$ is given by $ \var\{ f_g(\psi) \phi(\bZ; \Qb)\} \equiv f_g(\psi)^2 \sigma^2 $, where $f_g(\psi)$ is defined in Lemma \ref{lem:efficient influence function}, 
		\begin{align*}
			\sigma^2 &= \E\left( \frac{1}{\gamma(\bX)} \left[ \left\{ \frac{1-\gamma(\bX)}{\gamma(\bX)} \right\} \left\{ \frac{1-q_{12}(\bX)}{q_{12}(\bX)} \right\} + \frac{q_{0}(\bX) }{q_{12}(\bX)}\right] \right) + \var\left\{ \frac{1}{\gamma(\bX)} \right\}
		\end{align*}
		and $q_0(\bx) = 1 - q_1(\bx) - q_2(\bx) + q_{12}(\bx)$ is the chance of appearing on neither list 1 nor 2. 
	\end{theorem}
	
	\bigskip
	
	The magnitude of the efficiency bound in Theorem \ref{thm:var_phi} is driven by three main factors: 
	\begin{enumerate}
		\itemsep0em 
		\item[(i)] the magnitude of the conditional capture probabilities, 
		\item[(ii)] the chance of appearing on both lists, and
		\item[(iii)] the heterogeneity in the conditional capture probabilities.  \\
	\end{enumerate}
	
	\begin{remark}
		The efficiency bound for any function $g(\cdot)$ is always proportional to $\sigma^2$, with a scaling $f_g(\psi)^2$ depending on $g$; for example, $f_g(\psi)=-1$ when $g(\psi)=1/\psi$, and $f_g(\psi)=\psi/(1-\psi)$ when $g(\psi)=\logit(\psi)$. Therefore we focus our discussion on the quantity $\sigma^2$. \\
	\end{remark}
	
	The dependence on (i) in the bound in Theorem \ref{thm:var_phi} occurs through the term $(1-\gamma)/\gamma^2$, i.e., the odds of capture divided by the capture probability. The dependence on (ii) occurs through the odds $(1-q_{12})/q_{12}$ as well as the probability ratio $q_0/q_{12}$. The dependence on the heterogeneity (iii) occurs through the $\var(1/\gamma)$ term. Note that the probabilities $\gamma$ and $q_{12}$ in (i) and (ii) are related, but $q_{12}$ can be small even when the capture probability $\gamma$ is not, depending on the size of $q_1$ and $q_2$. \\
	
	More specifically, all else equal, the variance bound increases with: (i) smaller capture probabilities $\gamma$, (ii) smaller chances of appearing on both lists  $q_{12}$, and (iii) greater heterogeneity in the capture probabilities $\gamma$. Therefore capture probabilities can be estimated most efficiently when capture is likely, when there is substantial overlap across lists, and when capture probabilities are more homogeneous.  \\
	
	\begin{remark} For $K=2$ lists the quantity $q_0(\bx)= 0$ is exactly zero, but when $K>2$ it can be positive. \end{remark}
	
	\begin{remark}
		When $K=2$ and in the absence of covariates, the quantity $\sigma^2$ reduces to $ (\frac{1-\psi}{\psi^2})(\frac{1-q_{12}}{q_{12}})$. \\
	\end{remark}

	In addition to informing what factors yield more or less efficient capture probability estimation, the variance in Theorem \ref{thm:var_phi} also acts as a local minimax lower bound, as formalized in the following corollary. \\
	
	\begin{corollary}\label{cor:minimax}
		For any estimator $g(\widehat\psi)$, it follows that
		$$ \inf_{\delta>0} \ \liminf_{N \rightarrow \infty} \ \sup_{ \text{TV}(\overline\Qb, \Qb)<\delta } \ \frac{ \E_{\overline\Qb}\left[ \{ g(\widehat\psi) - g(\overline\psi) \}^2 \right] }{  f_g(\psi)^2 (\sigma^2 / N) } \geq 1 $$
		where $\text{TV}(\overline\Qb, \Qb)$ is the total variation between $\overline\Qb$ and $\Qb$, $\psi=\psi(\Qb)$ and $\overline\psi=\psi(\overline\Qb)$ are the capture probabilities under $\Qb$ and $\overline\Qb$, respectively, and $f_g(\psi)$ and $\sigma^2=\sigma^2(\Qb)$ are defined as in Theorem \ref{thm:var_phi}. 
	\end{corollary}
	
	Corollary \ref{cor:minimax} shows that the worst-case mean squared error of \emph{any} estimator of $\psi$, locally near the true $\Qb$, cannot be smaller than the efficiency bound, asymptotically and after scaling by $N$. This local minimax result gives an important benchmark for efficient estimation of the inverse capture probability: no estimator can have mean squared error uniformly better than the variance of the efficient influence function divided by $N$, without adding extra assumptions and/or structure to the nonparametric model we consider. \\
	
	\begin{remark}
		The local minimax result in Corollary \ref{cor:minimax} holds for any subconvex loss function $\ell: \R \mapsto [0,\infty)$ applied to $ \sqrt{N}\{g(\widehat\psi)-g(\overline\psi)\}$, not just squared error loss $\ell(t)=t^2$; the denominator lower bound in the general case is $\E\{ \ell(f_g \sigma Z)\} $ where $Z \sim \mathcal{N}(0,1)$ is a standard normal random variable \citep{van2002semiparametric}. \\
	\end{remark} 
	
	Importantly, in the next section we construct estimators that can achieve the nonparametric efficiency bound under weak conditions that allow for flexible estimation of the $q$-probabilities, e.g., using machine learning tools. \\
	
	\section{Efficient Estimation}
	\label{sec:estim}
	
	\subsection{Setup}
	
	Recall we let $\Qn$ denote the empirical measure under $\Qb$, so that sample averages can be written with the short-hand $\Qn(f)=\Qn\{f(\bZ)\}=\frac{1}{N} \sum_{i=1}^N f(\bZ_i)$. The simplest estimator of the capture probability $\psi$ is just a plug-in 
	\begin{equation} \label{eq:plugin}
		\widehat\psi_{pi} = \left[ \Qn \left\{ \frac{1}{\widehat\gamma(\bX)} \right\} \right]^{-1} 
	\end{equation}
	which replaces unknown quantities in the definition of $\psi$ with estimates, i.e., by estimating the conditional capture probability $\widehat{\gamma}(\bX) = \frac{\widehat{q}_{12} (\bX)}{\widehat{q}_1(\bX) \widehat{q}_2(\bX)}$ for every unit, and computing the harmonic mean of the values across the sample. This estimator has been used relatively extensively in previous work \citep{huggins2007non, Darroch1958indeplists, Fienberg1972loglinear, tilling1999capture}. 
	When the $q$-probabilities are estimated with correctly specified parametric models, the plug-in estimator $\widehat\psi_{pi}$ will be $\sqrt{n}$-consistent and asymptotically normal under standard regularity conditions. However, when the covariates contain any continuous components and/or are high-dimensional, it is usually very unlikely an analyst would have enough a priori knowledge to be able to correctly specify a low-dimensional parametric model, let alone three (one for each $q$-probability nuisance function). \\
	
	This difficulty of correct model specification suggests trying to flexibly estimate the $q$-probabilities, e.g., using logistic regression with model selection, or lasso, or nonparametric tools like random forests, neural nets, RKHS regression, etc. Unfortunately, when the plug-in estimator $\widehat\psi_{pi}$ is constructed from these kinds of data-adaptive methods, it in general loses the nice properties it has in the parametric setup. Specifically, without special tuning of particular methods, it in general would suffer from slower than $\sqrt{n}$-convergence rates, and have an unknown limiting distribution, making it not only inefficient but also leaving no tractable way to do inference. This deficiency of plug-in estimators is by now relatively well-known in functional estimation problems \citep{van2003unified, chernozhukov2018double, wu2019chebyshev}; however we have not seen it highlighted in the capture-recapture setting. (We show some of these issues numerically with simulations in Section \ref{sec:simulation}.) \\
	
	Luckily, the plug-in can be improved upon using tools from semiparametric efficiency theory \citep{bickel1993efficient, van2002part, van2003unified, tsiatis2006semiparametric, kennedy2016semiparametric}. In what follows, we will present and study a novel doubly robust estimator, which can attain the efficiency bound from the previous section even when built from flexible data-adaptive regression tools. \\
	
	\subsection{Doubly Robust Estimator}
	\label{sec:bias_corr}
	
	As mentioned above, the plug-in estimator \eqref{eq:plugin} has some important deficiencies in semi- and non-parametric settings. The plug-in estimator can be debiased by adding an estimate of the mean of the efficient influence function \citep{bickel1993efficient, van2002part, van2003unified, tsiatis2006semiparametric, kennedy2016semiparametric}. This leads to our proposed doubly robust estimator
	\begin{equation}\label{eq:biascorrection}
		\widehat\psi_{dr} = \Qn\left[ \frac{1}{\widehat{\gamma}(\bX)}\left\{\frac{Y_1}{\widehat{q}_{1}(\bX)} + \frac{Y_2}{\widehat{q}_{2}(\bX)} - \frac{Y_1 Y_2}{\widehat{q}_{12}(\bX)}\right\} \right]^{-1}
	\end{equation}
	where $\widehat{q}_j$ are estimates of the $q$-probabilities (e.g., via regression predictions). \\
	
	\begin{remark}
		In order to avoid potentially restrictive empirical process conditions, we estimate $\Qn$ and $\widehat{q}_j$ from separate independent samples. Specifically, we estimate the $q$-probability nuisance functions by fitting regressions in a training sample, independent of a test sample $\Qn$. With iid data, one can always obtain such samples by splitting at random in half, or folds. This yields a loss in efficiency, but that can be fixed by swapping the samples/folds, computing the estimate on each, and averaging. This is referred to as cross-fitting, and has been used for example by \citet{bickel1988estimating, robins2008higher, zheng2010asymptotic, chernozhukov2017double}.
		Here we analyze a single split procedure, merely to simplify notation; extending to averages across independent splits is straightforward. \\
	\end{remark}
	
	In the following sub-section, we derive finite-sample error bounds and distributional approximations for our doubly robust estimator, which are valid for any sample size. \\
	
	\subsubsection{Non-asymptotic Error Bounds and Approximate Normality}\label{sec:approx_norm}
	
	In this section, we provide our three main theoretical results regarding error bounds for our proposed method. In particular we show that our estimator is nearly efficient, doubly robust, and approximately normal. Importantly, we show all these properties hold in finite samples, without resorting to asymptotics. \\
	
	In the previous section, we derived the efficient influence function, which is the crucial component of the local minimax lower bound we gave in Corollary \ref{cor:minimax}. This corollary shows the minimax optimal estimator has mean squared error that scales like the variance of the efficient influence function divided by $N$, so that an optimal estimator would be one that behaves like a sample average of the efficient influence function. Our first result shows that our proposed estimator \emph{does} in fact behave like an average of the efficient influence function, depending on the size of nuisance error. In what follows, we use $\phi$ and $\widehat\phi$ to denote the efficient influence function $\phi(\bZ; \Qb)$ and its estimate $\phi(\bZ; \widehat\Qb)$ respectively. \\
	
	\begin{theorem}\label{thm:bc_errorterm}
		For any sample size $N$ and error tolerance $\delta>0$, we have
		$$ | (\widehat\psi^{-1}_{dr} - \psi^{-1}) - \Qn \phi | \leq \delta $$
		with probability at least
		$$ 1 - \left( \frac{1}{\delta^2} \right) \E\left( \widehat{R}_2^2  + \frac{ \| \widehat\phi - \phi \|^2 }{N} \right) $$
		where $\widehat{R}_2$ is a second-order error term given by
		\begin{align*}
			\widehat{R}_2 &= \int \frac{1}{\widehat{q}_{12}} \left\{ \Big( q_1 - \widehat{q}_1 \Big) \Big( \widehat{q}_2 - q_2 \Big) + \Big( q_{12} - \widehat{q}_{12} \Big) \left( \frac{1}{\gamma} - \frac{1}{\widehat\gamma} \right) \right\} \ d\Qb \\
			& \leq \left( \frac{1}{\epsilon} \right) \| \widehat{q}_1 - q_1 \| \| \widehat{q}_2 - q_2 \| + \left( \frac{1}{\epsilon^3} \right) \| \widehat{q}_{12} - q_{12} \| \| \widehat\gamma - \gamma \| 
		\end{align*}
		with the latter bound on $\widehat{R}_2$ holding as long as $(q_{12} \wedge \widehat{q}_{12}) \geq \epsilon$. 
	\end{theorem}
	
	Theorem \ref{thm:bc_errorterm} shows that our proposed estimator is within $\delta$ of a sample average of the efficient influence function, centered at the true (inverse) capture probability, with high probability, at every sample size. For a given observed number of captures $N$ and error $\delta$, this probability depends on two factors: (i) a second-order error term $\widehat{R}_2$, which is driven by the error in estimating the nuisance $q$-probabilities; and (ii) the $L_2$ error in estimating the efficient influence function itself, divided by $N$, which also depends on estimation error of the $q$-probabilities, but in a weaker way due to the division by $N$. When $\widehat{R}_2$ goes to 0 as $N$ increases, the probability above goes to 1 for any fixed $\delta$. Hence, Theorem \ref{thm:bc_errorterm} implies usual asymptotic convergence in probability, but in addition it gives an error bound that is valid for any finite $N$.  \\
	
	For example, if the $q$-probabilities are estimated with errors upper bounded by $cN^{-1/4}$, then with at least 95\% probability, our proposed estimator will be within $4c^2\sqrt{5/N}$ of the average efficient influence function. More generally, if $\E|\widehat{R}_2| \lesssim 1/\sqrt{N}$, then our proposed estimator will be within $1/\sqrt{N}$ (up to constants) of this efficient average, with high probability. For example, if $q_{1},\, q_{2}$, $q_{12}$ and $\gamma$ belong to Holder classes $\mathcal{H}(\beta_{1})$, $\mathcal{H}(\beta_{2})$, $\mathcal{H}(\beta_{12})$ and $\mathcal{H}(\beta_{0})$ respectively, where $\mathcal{H}(s)$ is a Holder class with smoothness index $s$ \citep{gyorfi2006distribution, tsybakov2008introduction}, then a sufficient condition for this kind of result, if the $q$-probabilities are estimated at minimax optimal rates, would be that the minimum smoothness is at least half the dimension of the covariates, i.e., $min\lbrace \beta_0, \beta_{12}, \beta_{1}, \beta_{2}\rbrace \ge d/2$. Similarly, if the $q$-probabilities were $s$-sparse, then a sufficient condition would be that $s \lesssim \sqrt{n}$ with lasso-style methods \citep{farrell2015robust}. 
	However, such $N^{-1/4}$ nuisance errors are only sufficient conditions for $1/\sqrt{N}$ capture probability errors; one only needs the remainder error $\widehat{R}_2$ to be small enough, which could also be achieved if some combinations of $q$-probabilities are estimated well, even if others are not. We give more detail on this phenomenon in our next result. \\
	
	Namely, beyond being close to an optimally efficient estimator, in the next result we show that our proposed estimator enjoys a finite-sample multiple robustness phenomenon, never before shown in capture-recapture problems. This phenomenon indicates that the overall estimation error can be small as long as some, but not all, nuisance probabilities are estimated with small error. \\
	
	\begin{corollary}\label{cor:doublerobust}
		Suppose $(q_{12}\wedge \widehat{q}_{12}) \geq \epsilon$. Assume one of the following holds: 
		\begin{enumerate}
			\item $\Vert \widehat{q}_{1} - q_{1}\Vert \vee \Vert \widehat{q}_{12} - q_{12}\Vert \leq \xi_N$, or
			\item $\Vert \widehat{q}_{1} - q_{1}\Vert \vee \Vert \widehat{\gamma} - \gamma\Vert \leq \xi_N$, or
			\item $\Vert \widehat{q}_{2} - q_{2}\Vert \vee \Vert \widehat{q}_{12} - q_{12}\Vert \leq \xi_N$, or
			\item $\Vert \widehat{q}_{2} - q_{2}\Vert \vee \Vert \widehat{\gamma} - \gamma\Vert \leq \xi_N$.
		\end{enumerate} 
		Then $ | (\widehat\psi^{-1}_{dr} - \psi^{-1}) - \Qn \phi | \leq \delta$ with probability at least
		$$ 1 - \left( \frac{C}{\delta^2} \right) \left( \xi_N^2 + \frac{1}{N} \right) $$
		where $C$ is a constant independent of sample size $N$.
	\end{corollary}
	
	As a consequence of Corollary \ref{cor:doublerobust}, the proposed estimator is doubly robust, i.e., if either estimator of $q_1$ or $q_2$ has small error, and either estimator of $q_{12}$ and $\gamma$ has small error, then the overall error of our proposed estimator (given by $\widehat{R}_2$) will be just as small, up to constants, even if the other estimators have large errors or are misspecified. (We note that, although this kind of robustness is sometimes called multiple robustness \citep{vansteelandt2008multiply}, we use the term doubly robust since the error structure is still second-order, i.e., involving products of errors, albeit with more terms). This property is very useful when one of the lists is difficult to estimate, for example, due to high-dimensional covariates, or $q$-probabilities that are complex functions of continuous covariates. 
	
	\begin{remark}
		Note that when there are only $K=2$ lists, then we have the relation $q_1(\bx)+q_2(\bx)-q_{12}(\bx)=1$ for all $\bx$. Hence, small bias for any two estimators of $q_{12}$, $q_{1}$ and $q_{2}$ automatically implies small bias for the third. One might expect then that double robustness does not arise in the $K=2$ list setting; however this is not quite true. To see why, note that it could be possible to estimate $\gamma$ with small error, for example, even when some of the $q$-probability estimators are misspecified. These and other issues related to estimation of the conditional capture probability $\gamma$ will be important to explore in future work.  \\
	\end{remark}
	
	When the remainder error $\widehat{R}_2$ is sufficiently small, Theorem \ref{thm:bc_errorterm} and Corollary \ref{cor:doublerobust} tell us we can approximate $\widehat\psi_{dr}^{-1}$ with a sample average of the efficient influence function. 
	For the purposes of inference, this suggests a confidence interval of the form
	\begin{equation} \label{eq:ci}
		\widehat{\text{CI}} = [\widehat\psi_{dr}^{-1} \pm z_{1-\alpha/2} \widehat\sigma/\sqrt{N} ],
	\end{equation}
	where $\widehat\sigma$ is a variance term defined in Theorem \ref{thm:normality}.
	In the next Berry-Esseen-type result, we exploit this closeness with a sample average and further show that our proposed estimator, properly scaled, is approximately Gaussian. This will show that the above confidence interval gives nearly-valid finite-sample coverage guarantees. \\
	
	\begin{theorem}\label{thm:normality}
		Let $\widehat\sigma^2 = \widehat{\var}(\widehat\phi)$ be the unbiased empirical variance of the estimated efficient influence function. Then $\widehat\psi^{-1}_{dr} - \psi^{-1}$ follows an approximately Gaussian distribution,  with the difference in cumulative distribution functions uniformly bounded above by
		\begin{align}
			\left\vert\Pb\left(\frac{\widehat\psi^{-1}_{dr} - \psi^{-1}}{\widehat\sigma/\sqrt{N}} 
			\le t\right) - \Phi(t)\right\vert &\leq \frac{C}{\sqrt{N}} \E\left( \frac{\rho}{\widetilde\sigma^3 }\right) + \frac{1}{\sqrt{2\pi}} \left\{ \sqrt{N}\E\left(  \frac{ |\widehat{R}_2|}{ \widetilde\sigma} \right) + |t| \E\left(  \left| \frac{\widehat\sigma}{\widetilde\sigma} - 1  \right| \right) \right\}\label{eq:approxnorm}
		\end{align}
		where $\widetilde\sigma = \var(\widehat\phi| \bZ^n)$, $\rho = \E\{|\widehat\phi - \Qb\widehat\phi|^3 \big| \bZ^n\}$ and $C<1/2$ is the Berry-Esseen constant.
	\end{theorem}
	
	\bigskip
	
	The above result shows that the estimation error scaled by $\widehat\sigma/\sqrt{N}$ is approximately standard normal. The first term on the right hand side of \eqref{eq:approxnorm} is the usual Berry-Esseen bound. The second term captures the effect of the nuisance estimation error $\widehat{R}_2$. The third term is the estimation error in the variance. Since $\E|\widehat\sigma - \widetilde\sigma|$ is bounded above by $cN^{-1/2}$ (proof in the appendix), the overall error in the Gaussian approximation is driven by the second term, involving nuisance error $\widehat{R}_2$. This will be the main driver of whether the interval has approximately correct coverage. We note that the above theorem implies convergence in distribution whenever $\E|\widehat{R}_2| = o(1/\sqrt{N})$ (which can hold for a wide variety of flexible nonparametric estimators of the $q$-probabilities, as discussed after Theorem \ref{thm:bc_errorterm}), but in addition gives a more precise error bound that holds for any finite sample size.\\
	
	Note that Theorem \ref{thm:normality} immediately implies that the error in coverage 
	$$ \left|\Pb\left( \widehat{\text{CI}} \ni \psi^{-1} \right) - (1-\alpha)\right| $$
	for the proposed confidence interval defined in \eqref{eq:ci} is no more than twice the error bound on the right-hand-side of \eqref{eq:approxnorm}, with $t=z_{\alpha/2}$. Further, a Berry-Esseen-style bound similar to that of Theorem \ref{thm:normality} (along with subsequent coverage guarantees and corollaries) can be obtained for any function $g(\cdot)$ of $\widehat\psi_{dr}^{-1}$ satisfying the conditions from \citet{friedrich1989berry}. This implies the same kind of coverage guarantees for $\psi$, for example, using the confidence interval
	$$ \widehat\psi_{dr} \pm z_{1-\alpha/2} \widehat\sigma\widehat\psi_{dr}^2 /\sqrt{N} $$
	which can be motivated via the delta method. The error in the coverage of this estimated interval is twice the bound in Theorem \ref{thm:normality}, modulo some extra dependence on $g$.
	
	Importantly, the unbiased empirical variance $\widehat\sigma^2$ is a consistent estimator of the efficiency bound $\sigma^2= \var(\phi)$ in the sense that $\E|\sigma - \widehat\sigma| \lesssim \E\|\widehat\phi - \phi\| + N^{-1/2}$. This shows the crucial result that our estimator is approximately minimax optimal in the sense of Corollary \ref{cor:minimax}, if the nuisance error is small enough.  \\
	
	A natural consequence of the above theorem is the following corollary, which presents a simple bound on the error of the normal approximation, under some natural conditions on the nuisance error $\widehat{R}_2$ and variance. \\
	
	\begin{corollary}\label{cor:asymp_psi}
		Assume $\widetilde\sigma \gtrsim 1$, $\E| \widehat{R}_2| \lesssim N^{-2\beta}$ and $\alpha > \delta$ for some $\delta>0$. Then the coverage error for the proposed $(1-\alpha)$ confidence interval defined in \eqref{eq:ci} is upper bounded by 
		$$ \left|\Pb\left( \widehat{\text{CI}} \ni \psi^{-1} \right) - (1-\alpha)\right| \ \lesssim \ {N^{(1-4\beta)/2}} + \frac{1}{\sqrt{N}} . $$
		Therefore if $\beta>1/4$ there exists some sample size $N_\epsilon$ at which the coverage error is never more than $\epsilon$, for any $N>N_\epsilon$. \\
	\end{corollary}
	
	Since this corollary is a special case of Theorem \ref{thm:normality}, mainly aimed at presenting the result in a simple form, we refer to the above discussion for more details. However we note that the condition that $\E|\widehat{R}_2| \lesssim N^{-2\beta}$ would hold for example if the $q$-probabilities were estimated optimally when contained in Holder classes with smoothness index $s$, where $\beta=\frac{s}{2s+d}$ (or under some conditions on sparsity, as discussed after Theorem \ref{thm:bc_errorterm}). Then $\beta > 1/4$ would mean $s>d/2$, aligning with our earlier results. \\
	
	In this section, we have given finite-sample error bounds and distributional approximations for our proposed estimator, which are valid for any sample size, allowing accurate estimation and approximately valid confidence guarantees, even in complex nonparametric models where the $q$-probabilities are estimated with flexible machine learning tools. In the next section, we consider a slightly modified version of the estimator which could further improve finite-sample properties.  \\
	
	\subsection{Targeted Maximum Likelihood Estimator}
	\label{sec:tmle}
	
	We have seen that our proposed doubly robust estimator \eqref{eq:biascorrection} is close in a finite-sample sense to an optimal sample average, and possesses crucial double robustness properties. However it is possible this estimator may not respect the bounds on the parameter space; for example $\widehat\psi_{dr}$ may fall outside $[0,1]$ if some of the estimates of the $q$-probabilities are small. A simple fix is to truncate the estimator $\widehat\psi_{dr}$ to always lie in $[0, 1]$. Here for completeness we discuss an alternative approach using targeted maximum likelihood estimation (TMLE) \citep{van2006targeted, van2011targeted}, which is an iterative procedure that fluctuates nuisance estimates so that a plug-in estimator built from them also approximately solves an efficient influence function estimating equation. TMLE thus leads to estimators that are asymptotically equivalent to one-step bias-corrected estimators, but which could bring some finite-sample advantages. \\
	
	Below we present an algorithm detailing the computation of a TMLE for $\psi$. At a high level, the procedure involves iterative updating of initial nuisance estimates, based on the quantities $\bH_{12,t},\, \bH_{1,t}$, and $\bH_{2,t}$ defined shortly, which are called clever covariates in the TMLE literature.
	
	\begin{algorithm}TMLE algorithm for estimating $\psi$\label{algo:tmle}
		\begin{enumerate}
			\item Obtain initial estimates of $q_{12}(\bx)$, $q_1(\bx)$ and $q_2(\bx)$, denoted $\widehat{q}_{12,0}(\bx)$, $\widehat{q}_{1,0}(\bx)$ and $\widehat{q}_{2,0}(\bx)$. Set $t =0$.
			\item At step $t$, construct clever covariates: \begin{enumerate}\item $\bH_{12,t} =   \frac{\widehat{q}_{1,t}(\bX)\widehat{q}_{2,t}(\bX)}{\widehat{q}_{12,t}(\bX)^2} - \frac{\widehat{q}_{1,t}(\bX)}{\widehat{q}_{12,t}(\bX)} - \frac{\widehat{q}_{2,t}(\bX)}{\widehat{q}_{12,t}(\bX)}$
				\item $\bH_{1,t} = \frac{\widehat{q}_{2,t}(\bX)}{\widehat{q}_{12,t}(\bX)}$
				\item $\bH_{2,t} = \frac{\widehat{q}_{1,t}(\bX)}{\widehat{q}_{12,t}(\bX)}$.
			\end{enumerate} 
			\item Regress $Y_1Y_2$ on $\bH_{12,t}$ using a no-intercept logistic model with $\logit\{\widehat{q}_{12,t}(\bX)\}$ as offset, obtaining estimated coefficient $\widehat\beta_{12,t}$.
			Set $\widehat{q}_{12,t+1}(\bX) = \expit\left[ \logit\{\widehat{q}_{12,t}(\bX)\} + \widehat{\beta}_{12} \bH_{12,t}\right]$.
			\item Regress $Y_1(1 - Y_2)$ on $\bH_{1,t}$ using a no-intercept logistic model with $\logit\{\widehat{q}_{1,t}(\bX) - \widehat{q}_{12, t+1}(\bX)\}$ as offset, obtaining estimated coefficient $\widehat\beta_{1,t}$.
			Set $$\widehat{q}_{1,t+1}(\bX) = min \left\{ \widehat{q}_{12, t+1}(\bX) + \expit\left[ \logit\{\widehat{q}_{1,t}(\bX) - \widehat{q}_{12,t+1}(\bX)\} + \widehat{\beta}_{1,t} \bH_{1,t}\right], 1 - \widehat{q}_{12, t+1}(\bX)\right\}. $$
			\item Regress $Y_2(1 - Y_1)$ on $\bH_{2,t}$ using a no-intercept logistic model with $\logit\{\widehat{q}_{2,t}(\bx) - \widehat{q}_{12,t+1}(\bx)\}$ as offset, obtaining estimated coefficient $\widehat\beta_{2,t}$.
			Set 
			\begin{align*}
				\widehat{q}_{2,t+1}(\bX) &= min \left\{\widehat{q}_{12, t+1}(\bX) + \expit\left[ \logit\{\widehat{q}_{2,t}(\bX) - \widehat{q}_{12,t+1}(\bX)\} + \widehat{\beta}_{2,t} \bH_{2,t}\right],\right.\\
				& \hspace{.5in} 
				\Big.1 + \widehat{q}_{12,t+1}(\bX) - \widehat{q}_{1,t+1}(\bX)\Big\} .     
			\end{align*}
			\item Update $t \longrightarrow t + 1$. Repeat Steps 2 to 6 until convergence (e.g., until $\max_j |\widehat\beta_{j,t+1}| \leq \epsilon$).
		\end{enumerate}
		Finally, set $\widehat\psi_{tmle} = \left[\Qn\left\{\frac{\widehat{q}^*_{1}(\bX)\widehat{q}^*_{2}(\bX)}{\widehat{q}^*_{12}(\bX)}\right\}\right]^{-1}$, where $\widehat{q}_j^*$ are estimates obtained after convergence.
	\end{algorithm}
	
	\bigskip
	
	\begin{remark}
		Step 5 in the algorithm can be modified so that for $K = 2$, $q_2$ is evaluated by $\widehat{q}_{2,t+1}(\bx) = 1 + \widehat{q}_{12, t+1} - \widehat{q}_{1,t+1}(\bx)$. This step uses the relation that for $K=2$, $q_1 + q_2 - q_{12} = 1$. \\
	\end{remark}
	
	Interestingly, since the TMLE above is not a sample average like our main proposed estimator from the previous subsection, it is less clear how to derive finite-sample error bounds, despite this being the main motivation for TMLE. Since the 
	estimates $\widehat{q}^*_{j}(\bx)$ obtained after convergence satisfy $\Qn\{\phi(\bZ; \widehat{\Qb}^*)\} \approx 0$, the asymptotic behavior matches the doubly robust estimator in \eqref{eq:biascorrection}, but for describing finite-sample behavior we resort to simulations, detailed in Section \ref{sec:simulation}. \\ 
	
	\section{Inference for Population Size}\label{sec:population_size}
	
	In the previous section we gave doubly robust estimators for the capture probability and studied finite-sample properties. In this section we give a crucial result that shows how to obtain an approximate confidence interval for the \emph{population size}, given a generic initial estimator of the (inverse) capture probability. Importantly, our results only require this initial estimator to be weakly approximated by a sample average, and otherwise are completely agnostic to how the capture probability is estimated. This appears in stark contrast to most of the literature on this topic, where the inferential procedures are very closely tied to specific model assumptions and estimator constructions. \\
	
	This main inferential result is given in the following theorem.
	
	\begin{theorem}\label{thm:confidenceinterval}
		Suppose we are given an initial estimator $\widehat\psi$ that satisfies
		$$\widehat\psi^{-1} - \psi^{-1} = \Qn\left(\widehat\varphi\right) - \int\widehat\varphi(\bz)d\Qb(\bz) + \widehat{R}_2$$
		for $\varphi$ a generic influence function with mean zero and $\widehat{R}_2$ an error term. Let $\widehat\tau^2 = \widehat\psi\widehat\varsigma^2 + \frac{1 - \widehat\psi}{\widehat\psi}$ and $\widetilde\tau^2 = \psi\widetilde\varsigma^2 + \frac{1 - \psi}{\psi}(\psi\widehat{R}_2 + 1)^2$, where
		$\widehat\varsigma^2 = \widehat{\var}(\widehat\varphi)$ is the unbiased empirical variance of the estimated influence function and $\widetilde\varsigma = \var(\widehat\varphi \mid \bZ^n)$ the true conditional variance. 
		Then the $(1-\alpha)$ confidence interval given by 
		\begin{align}
			\widehat{\text{CI}}_n = \left[\widehat{n} \pm z_{\alpha/2} \widehat\tau\sqrt{\widehat{n}} \right]\label{eq:cin}
		\end{align}
		has coverage error upper bounded as
		\begin{align}
			\left|\Pb\left( \widehat{\text{CI}}_n \ni n \right) - (1-\alpha)\right| &\leq \frac{2C}{\sqrt{n}} \E\left( \frac{\rho}{\widetilde\tau^3} \right) + \sqrt{\frac{2}{\pi}} \left\{ \sqrt{n} \psi\E\left(\frac{|\widehat{R}_2|}{\widetilde\tau}\right) + |z_{\alpha/2}| \E\left(  \left|\frac{\widehat\tau\sqrt{\widehat{n}}}{\widetilde\tau\sqrt{n}}- 1  \right| \right) \right\}\label{eq:normal_error}
		\end{align}
		where $C$ is the Berry-Esseen constant and
		$$\rho = \E\left[\left|\one(\bY \neq \bzero)\left\{\widehat{\varphi} - \Qb\widehat{\varphi}\right\} + \left\{\one(\bY \neq \bzero) - \psi\right\}\widehat{R}_2 + \psi^{-1}\left\{\one(\bY \neq \bzero) - \psi\right\}\right|^3 \Bigg\vert\bZ^n\right].$$
	\end{theorem}
	
	\bigskip
	
	Theorem \ref{thm:confidenceinterval} gives a non-asymptotic upper bound on how much the coverage $\Pb( \widehat{\text{CI}}_n \ni n )$ of our proposed interval 
	$$ \frac{N}{\widehat\psi} \pm z_{\alpha/2} \sqrt{\left( \widehat\psi\widehat\varsigma^2 + \frac{1 - \widehat\psi}{\widehat\psi} \right) \frac{N}{\widehat\psi}} $$
	can deviate from its nominal $(1-\alpha)$ level. Before describing the coverage guarantee, we first describe the proposed interval. The length of this interval is driven by three factors: (i) the estimated odds of not being captured $(1-\widehat\psi)/\widehat\psi$, (ii) the variance of the inverse capture probability estimator $\widehat\varsigma^2$, and (iii) the sample size $N$. As one would expect, higher odds of capture yield more precise inference about population size, all else equal, as does more efficient estimation of $\widehat\psi$. Specifically, the length of the interval shrinks to zero when the capture probability is very large, regardless of the sample size $N$. Also note that even if $\psi$ were known, one would still have an interval of the form
	$$ \frac{N}{\psi} \pm z_{\alpha/2} \sqrt{\frac{1-\psi}{\psi} } \sqrt{ \frac{N}{\psi}} $$
	based on the fact that $\widehat{n} = N/\sqrt{n}$ is approximately normal. Although sample size $N$ appears in the numerator of the interval width (contrary to standard intervals), it only appears through its square root, showing that in an asymptotic regime where $N \rightarrow \infty$, the width still grows at a slower rate than the sample size. Intuitively, one can think of this interval as taking $\widehat{n} = N/\widehat\psi$ and multiplying by $1 \pm z_{\alpha/2}/\sqrt{\widehat{n}}$, which does tend to zero as sample size $N$ grows. \\
	
	\begin{remark}
		For $K = 2$ lists and in the absence of covariates, the confidence interval reduces to $\widehat{n} \pm z_{\alpha/2}\sqrt{\frac{\widehat{n} (1 - \widehat\psi)}{\widehat\psi \ \widehat{q}_{12}}}$, which approximately resembles the Wald-type confidence interval for the Petersen estimator  \citep{evans1996application}.\\
	\end{remark}
	
	Now we describe the coverage guarantee of Theorem \ref{thm:confidenceinterval}. Importantly, the bound on the coverage error depends on a number of factors, as shown above appearing the sum of the three terms in \eqref{eq:normal_error}. Under typical boundedness assumptions, the first and third terms would be of smaller order, and the second term would dominate. This second term is driven by the size of $\widehat{R}_2$ in terms of its mean absolute value, i.e., how well the initial estimator $\widehat\psi$ is approximated by a sample average. If $\widehat{R}_2$ is not substantially smaller than $1/\sqrt{n}$, then the confidence interval would not be guaranteed to cover the true population size $n$ at its nominal level. This points to the importance of efficient estimation of $\psi$; for example, as shown in the previous section, our proposed estimator $\widehat\psi_{dr}$ can be approximated by a sample average up to smaller than $1/\sqrt{n}$ error, even in a nonparametric model when $q$-probabilities are estimated flexibly.  \\
	
	\begin{remark}
		A unique feature of Theorem \ref{thm:confidenceinterval} is that it is valid for \emph{any} estimator approximated by a sample average, regardless of what underlying identification or estimation assumptions were used in its construction. This means if another analyst did not believe the independent lists condition in Assumption \ref{as:1}, and instead constructed an estimate of the capture probability under a different identifying assumption, they could also use the above theorem to construct a confidence interval and assess its finite-sample coverage. \\
	\end{remark}
	
	A natural consequence of Theorem \ref{thm:confidenceinterval} is the following corollary, which parallels Corollary \ref{cor:asymp_psi} in giving a simple bound on the error of the normal approximation, under some natural conditions. \\
	
	\begin{corollary}\label{cor:asymp_n}
		Assume $\widetilde\tau \gtrsim 1$, $\E| \widehat{R}_2| \lesssim N^{-2\beta}$ and $\alpha > \delta$ for some $\delta>0$. Then the coverage error for the proposed $(1-\alpha)$ confidence interval defined in \eqref{eq:ci} is upper bounded by 
		$$ \left|\Pb\left( \widehat{\text{CI}_n} \ni n \right) - (1-\alpha)\right| \ \lesssim \ {n^{(1-4\beta)/2}} + \frac{1}{\sqrt{n}} . $$
		Therefore if $\beta>1/4$ there exists some population size $n_\epsilon$ at which the coverage error is never more than $\epsilon$, for any $n>n_\epsilon$. \\
	\end{corollary}
	
	Since the result in Corollary \ref{cor:asymp_n} is similar to that of Corollary \ref{cor:asymp_psi}, we refer there for related discussion. The main point is that, as long as our initial estimator is well-approximated by a sample average, no matter how it was constructed or what assumptions it relies on, our proposed confidence interval \eqref{eq:cin} will be approximately valid.\\
	
	\section{Simulation \& Application}
	\label{sec:results}
	
	So far we have proposed doubly robust estimators for the capture probability, and a general approach for constructing confidence intervals for the total population size, all with non-asymptotic error guarantees. In this section we study the performance of our methods in simulated data, and apply them to estimate the total number of killings in the internal armed conflict in Peru during 1980-2000. The code used to generate the results is available at \verb+mqnjqrid/capture_recapture+. \\
	
	\subsection{Simulation}\label{sec:simulation}
	
	Here we use a simulation set-up similar to that of \citet{tilling1999capture}. Specifically we take $n=5000$ samples from 
	\begin{align*}
		&X \sim \mathit{Uniform}(2, 3)\\
		&\Pb(Y_1 = 1 \mid X=x) = \expit(a + 0.4x)\\
		&\Pb(Y_2 = 1 \mid X=x) = \expit(a + 0.3x).
	\end{align*}
	where $a$ takes values $\{-2.513,\, -1.758,\, -0.66\}$ to ensure that the capture probability $\psi$ takes values $\lbrace 0.3,\, 0.5,\, 0.8 \rbrace$, respectively. This gives sample sizes $N$ approximately equal to $\{1500,2500,4000\}$. Recall that under $\Pb$, list membership $Y_1$ and $Y_2$ are conditionally independent, so the conditional capture probability is 
	$$ \gamma(x) = 1 - \{1 - \expit(a + 0.4x)\}\{1-\expit(a + 0.3x)\}$$
	and $q$-probabilities are equal to $q_j(x) = \Pb(Y_j=1 \mid X=x)/\gamma(x)$. \\
	
\begin{figure}[h!]
		\centering
		\includegraphics[scale = 0.62]{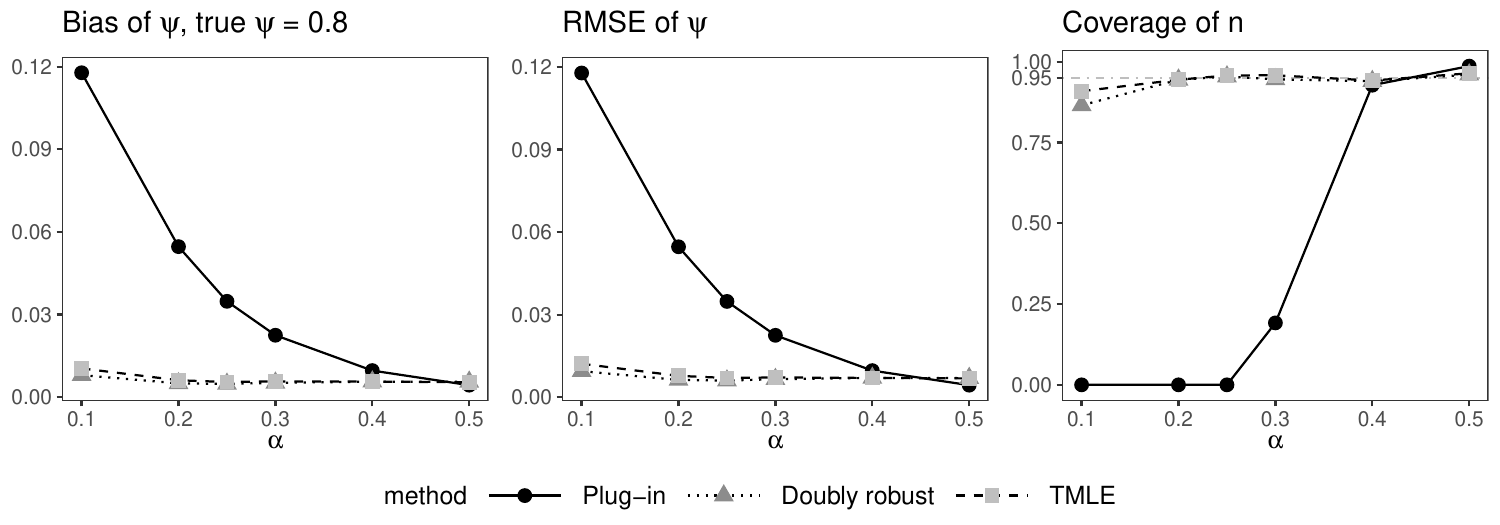}\vspace{-18pt}
		\includegraphics[scale = 0.62]{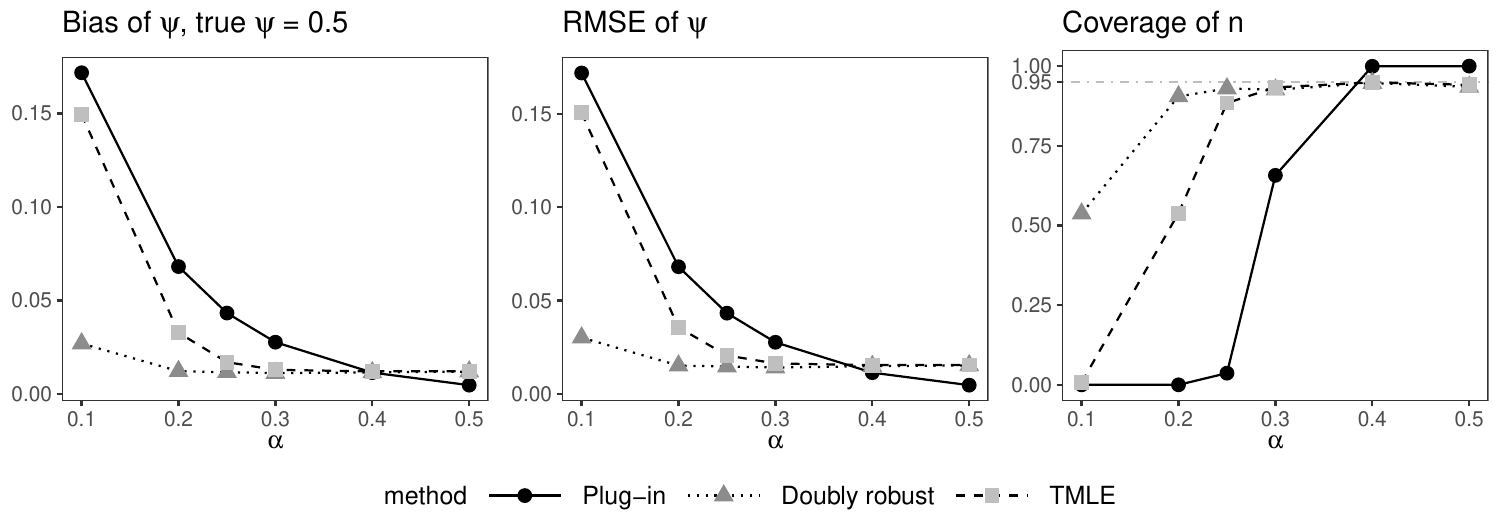}\vspace{-18pt}
		\includegraphics[scale = 0.62]{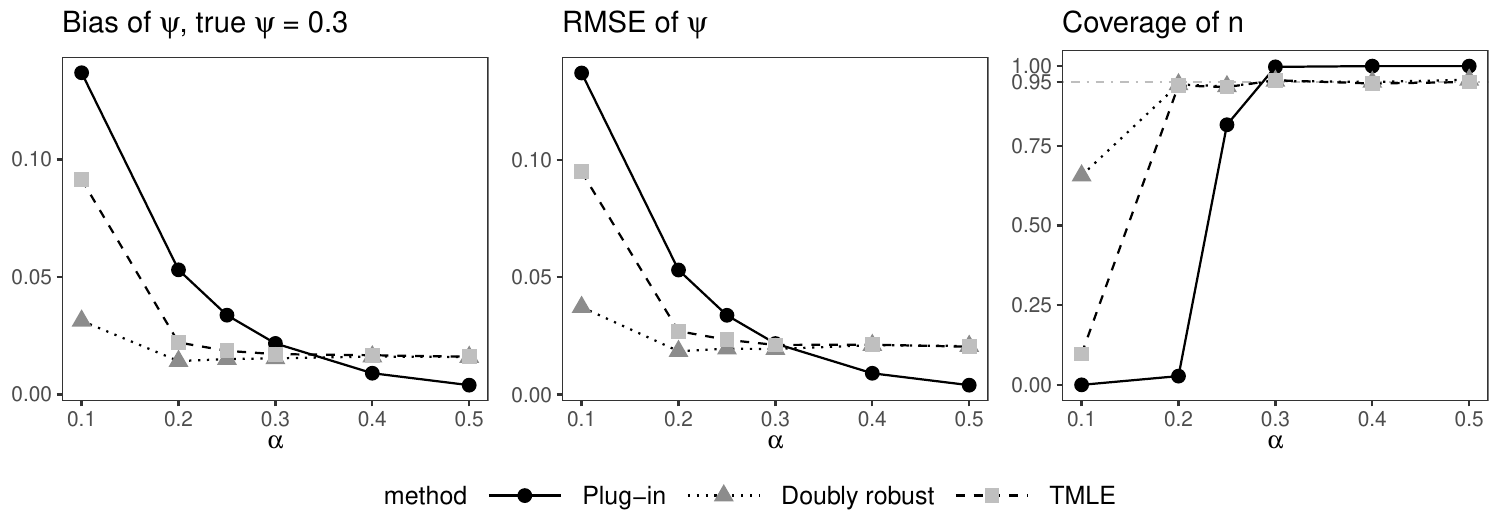}
		\caption{Estimated bias, RMSE, and population size coverage, for simulated data with population size $n=5000$, across true capture probability $\psi \in \{ 0.8 , 0.5 , 0.3\}$, $q$-probability error rate $n^{-\alpha}$ for $\alpha \in [0.1,0.5]$, and for three different estimators: the plug-in and two proposed doubly robust estimators.}
		\label{fig:barplot_diff_psi}
	\end{figure}

	We construct estimates of the $q$-probabilities via $\widehat{q}_j(x) = \expit[\logit\{q_j(x)\} + \epsilon_j ]$, where $\epsilon \sim \mathcal{N}(n^{-\alpha}, n^{-2\alpha})$. This allows us to carefully control the error of the $q$-probability estimators; since the root mean squared error scales like $n^{-\alpha}$, this can be viewed as the rate of convergence. We run 500 simulations for $\alpha \in \lbrace 0.1, 0.2, 0.25, 0.3, 0.4, 0.5\rbrace$. Note $\alpha$ values 0.5 and 0.25 correspond to the parametric ($n^{-1/2}$) and nonparametric ($n^{-1/4}$) rates, respectively. Figure \ref{fig:barplot_diff_psi} shows the estimated bias and the root mean square error (RMSE) of $\widehat\psi$, along with the coverage proportion for the confidence interval of the total population size. \\
	
	\begin{remark}
		For plug-in estimators, there is no well-defined variance formula (this is a main motivation for our doubly robust construction). Therefore to construct confidence intervals with the plug-in estimator, we used the estimated variance of the doubly robust estimator. \\
	\end{remark}
	
	Overall, the simulations illustrate the phenomena expected from our theoretical results: when the $q$-probabilities are estimated with low error (i.e., $\alpha$ large), all the methods do well, whereas when the $q$-probabilities are difficult to estimate (i.e., $\alpha$ smaller) the proposed methods do substantially better in terms of bias, error, and coverage. For example, when the true capture probability is 50\%, the simple plug-in estimator gives substantial bias as soon as $\alpha<0.4$ (i.e., when the $q$-probabilities are estimated at slower than $n^{-2/5}$ rates). However, the bias of the proposed doubly robust estimator is relatively unaffected until $\alpha< 0.2$, with the TMLE somewhere in between. The story is similar for the RMSE, which is largely driven by the bias in this problem. The coverage is approximately at the nominal 95\% level as soon as $\alpha>0.2$ (i.e., when the $q$-probabilities are estimated at faster than $n^{-1/5}$ rates), whereas the plug-in estimator substantially under-covers (e.g., nearly zero at $\alpha=0.2$) until $\alpha \geq 0.4$. We note that when population size or capture probability is small (e.g., capture probability substantially less than $50\%$), estimation becomes more challenging and the story is less clear about which method does better. For reference, results for population sizes varying from $n=5000$ to $n=25000$ (in the $\alpha = 0.25$ case) are given in the Appendix in Figure \ref{fig:lineplot_psi}. \\
	
	\subsection{Data Analysis}
	
	We apply our proposed methods to estimate the number of killings and disappearances attributable to different groups in Peru during its internal armed conflict between 1980 and 2000. We use data collected by the Truth and Reconciliation commission of Peru \citep{ball2003many}, as well as detailed geographic information, following \citet{rendon2019capturing}.\\
	
	There is an ongoing debate regarding the total number of killings and disappearances in the conflict, as well as about which groups are most responsible, e.g., the PCP-Shining Path versus the State or other groups. \citet{ball2003many} estimated approximately 69,000 total killings and disappearences, finding the Shining Path responsible for the majority. In contrast, \citet{rendon2019capturing} estimated approximately 48,000 killings and disappearances, with the State responsible for the majority, though many geographic strata were excluded. Most recently, \citet{manrique2019reality} included a newly available list and estimated approximately 58,000-65,000 killings and disappearances, depending on choices of priors, with the Shining Path responsible for the majority. \\
	
\begin{figure}[h!]
		\centering
		\includegraphics[scale=0.7]{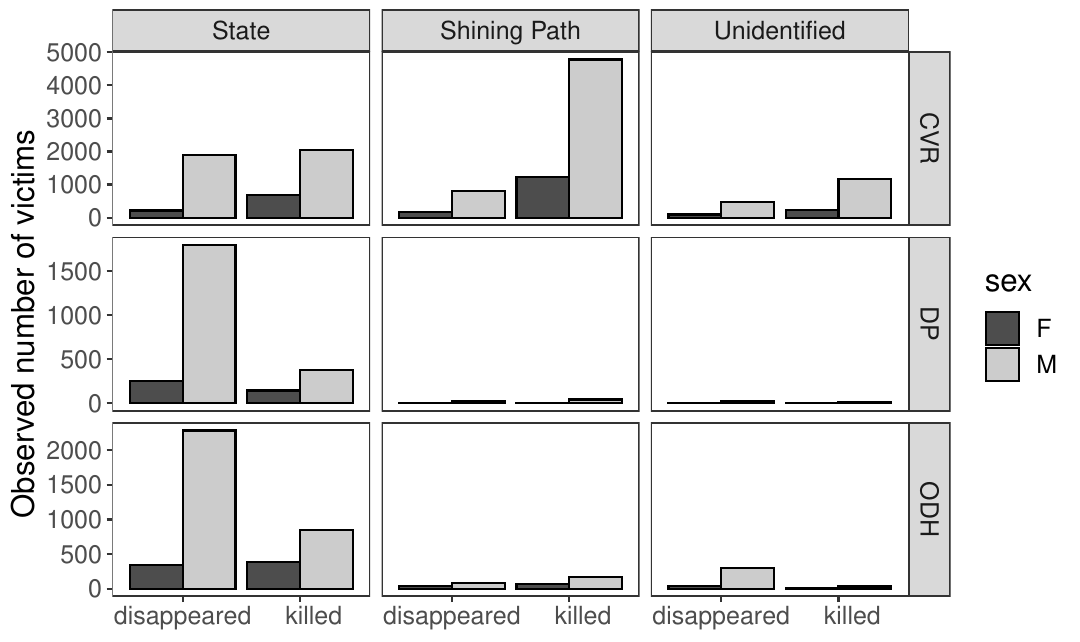}
		\caption{This figure shows the observed number of victims for the three lists (CVR, DP, ODH) across the State, the Shining Path and the victims with unidentified perpetrator. Most of the victims are males. The Truth and Reconciliation Commission (CVR) has documented the highest number of victims for the PCP-Shining Path compared to the defender of the People (DP) and the combined NGO's (ODH), whereas the later two sources documented most of the victims for the State and very few (<70 by DP and <400 by ODH) for the Shining Path. Majority of the victims of the State disappeared whereas, most of the victims of the Shining Path were killed.}
		\label{fig:demo_covariate}
	\end{figure}

\begin{figure}[h!]
		\centering
		\includegraphics[scale = 0.87]{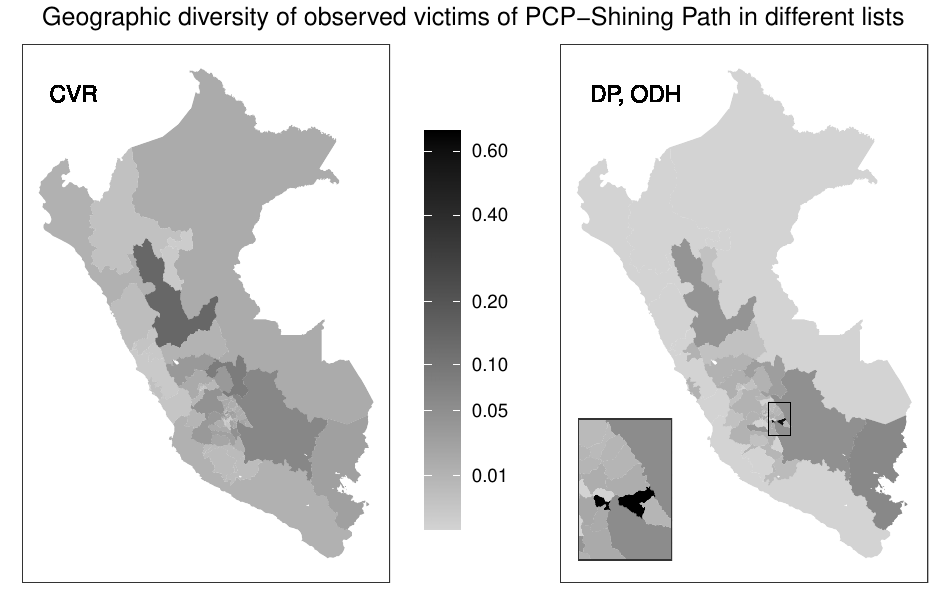}
		\caption{Geographic diversity of Shining Path victims at strata level, for CVR list (left) and DP and ODH lists (right). The color of each stratum reflects the proportion of all Shining Path victims in the list who were killed or disappeared in that stratum.}
		\label{fig:list_heterogeneity}
	\end{figure}

	Before describing our specific approach, we first describe the data and give some summary statistics. As explained in \citet{ball2003many}, the data come from a few main sources: the Truth and Reconciliation Commission (CVR), the Public Defender Office (DP), and 4--5 other human rights groups and NGOs (ODH); we combined the DP and ODH lists since they have similar demographics. The data contains identifiers of people who have been killed or disappeared, as well as which of the source lists they appeared on, and covariates including age, gender, and geographic location of the killing or disappearance (measured via 58 geographic strata as in \citet{ball2003many}, as well as bivariate latitude/longitude as in \citet{rendon2019capturing}). To avoid missing completely at random assumptions, we also included missingness indicators for victims with missing age (28\% missing), gender (<1\% missing), or location (11\% missing). The lists of all the covariates is available in the appendix \ref{app:realdata}. The total number of killings and disappearances across all lists was 24,692. Importantly, the lists capture different demographics, which points to the necessity of relaxing classical marginal independence via the \emph{conditional} independence in Assumption \ref{as:1}. For example, the CVR list mostly includes victims who were killed, while the DP and ODH lists mostly include victims who disappeared, as shown in Figure \ref{fig:demo_covariate}. Similarly, geographic diversity varies across lists, as shown in Figure \ref{fig:list_heterogeneity}. For example, almost 60\% of Shining Path victims in the DP and ODH lists come from two smaller districts (Chungui and Luis Carranzo) of Ayacucho, while in the CVR list the Shining Path victims are more uniformly spread across the country. More details on the data are available in Appendix \ref{app:realdata}. \\
	
	\bigskip
	
\begin{figure}[h!]
		\centering
		\includegraphics[scale = 0.7]{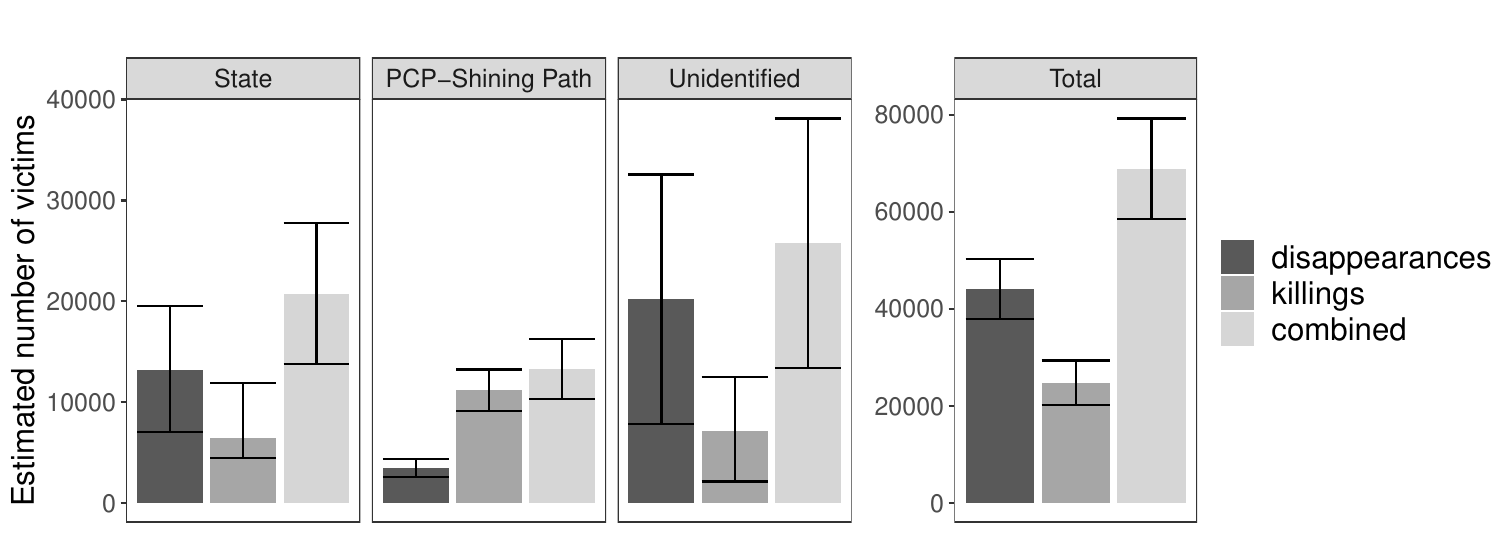}
		\caption{Estimated numbers of disappearances and killings (and both together) by perpetrator, as well as total (combined across perpetrators), using the proposed doubly robust method. Bars indicate 95\% confidence intervals.}\label{fig:perukill_barplot}
	\end{figure}

	Now we move to our analysis. Our goal was to estimate the number of killings and disappearances attributable to the State and Shining Path, as well as those that were not identified as either. We used our proposed doubly robust estimator \eqref{eq:biascorrection} with five-fold cross-fitting, and we estimated the $q$-probabilities via random forests (using the \verb|ranger| package in R). We truncated all $q$-probability estimates at 0.01. Figure \ref{fig:perukill_barplot} shows the estimated number of killings and disappearances along with 95\% confidence intervals obtained using the interval \eqref{eq:cin}. We estimate the total number of killings and disappearances across groups to be 68,874 (95\% CI: 58,543-79,204), close to the estimates in \citet{ball2003many} and the diffuse prior-based estimate in \citet{manrique2019reality} (which used an additional list). 
	Overall we find the State responsible for more disappearances, and Shining Path responsible for more killings; however we estimate the number of killings and disappearances by unidentified perpetrators to be larger than that for either group. In terms of the overall killings and disappearances, the estimate for the State are higher compared to the estimate for the Shining Path. We present some more details of the analysis and a location wise estimate comparison for the State and the Shining Path in Appendix \ref{app:realdata}. 
	
	\bigskip
	
	\section{Discussion}
	
	In this paper, we study estimation of population size and capture probability in the capture-recapture set-up where two lists are conditionally independent given measured covariates. We make four main contributions to the literature. First, we derive the nonparametric efficiency bound for estimating the capture probability, which indicates the best possible performance of any estimator, in a local asymptotic minimax sense. As far as we know this kind of lower bound result has not appeared in the literature, even in simple settings without covariates. Second, we present a new doubly robust estimator, and study its finite-sample properties; in addition to double robustness, we show that it is near-optimal in a non-asymptotic sense, under relatively mild nonparametric conditions. Third, we give a method for constructing confidence intervals for total population size from generic capture probability estimators, and prove non-asymptotic near-validity. And fourth, we study our methods in simulations, and apply them to estimate the number of killings and disappearances attributable to different groups in Peru during its internal armed conflict between 1980 and 2000. \\
	
	There are many ways one could extend and build on the work in this paper. For example, rather than assuming a known pair of lists are conditionally independent given the covariates, one could instead take a sensitivity analysis and/or partial identification approach. For example, one could assume that a pair of lists is only nearly conditionally independent, up to some deviation $\delta$, and estimate bounds on the capture probability and population size accordingly. This relies on weaker assumptions, with the trade-off of yielding less precise inferences. Another extension would be to flexibly estimate conditional capture probability or population size, given a continuous covariate such as age or time. For example, for the internal armed conflict in Peru it might be of interest to estimate the number of victims by age. This would require a non-trivial extension of the current methods, but would be important future work.
	
	\section*{Acknowledgements}
	
	The authors thank Mark van der Laan, and Robin Mejia for very helpful discussions. 
	Manjari Das and Edward Kennedy gratefully acknowledge support from the Berkman Faculty Development Fund, and Edward Kennedy from NSF Grant DMS1810979.

	\newpage
	\appendix
	
	\section{Proofs of theorems and results}\label{app:proofresult}
	
	\begin{proof}[\textbf{Proof of proposition \ref{thm:identification}}] The goal is to express $\psi$ in terms of the observed data distribution $\Qb$. By definition $\psi = \Pb(\bY \neq \bzero)$ and $\gamma(\bx) = \Pb(\bY \neq \bzero | \bX = \bx)$.\\[1.5ex]
		First we show that $\psi$ is the harmonic mean of $\gamma(\bx)$. It is easy to see that
		$$\frac{\gamma(\bx)}{\psi} = \frac{\Pb(\bY \neq \bzero | \bX = \bx)}{\Pb(\bY \neq \bzero)} = \frac{\Pb(\bY \neq \bzero,\, \bX = \bx)}{\Pb(\bX = \bx)\Pb(\bY \neq \bzero)} = \frac{\Pb(\bX = \bx | \bY \neq \bzero)}{\Pb(\bX = \bx)}.$$
		Thus,
		\begin{align*}
			&\int \frac{\Pb(\bX = \bx)}{\psi} d\bx = \int \frac{\Pb(\bX = \bx | \bY \neq \bzero)}{\gamma(\bx)} d\bx\\
			\text{or, }& \frac{1}{\psi} = \int\frac{\Qb(\bX = \bx)}{\gamma(\bx)} d\bx, \text{ since } \Qb(\bZ) = \Pb(\bZ | \bY \neq \bzero).
		\end{align*}
		We have shown a that the capture probability $\psi$ is the harmonic mean of the conditional capture probabilities $\gamma(\bx)$ under the observed data distribution $\Qb$.\\[1.5ex]
		
		Next, we express $\gamma(\bx)$ in terms of the $q$-probabilities. Under assumption \ref{as:1}, for any fixed covariate value $\bx$,
		$$\Pb(Y_1 = 1,\, Y_2 = 1|\bX = \bx) = \Pb(Y_1 = 1|\bX = \bx)\Pb(Y_2 = 1|\bX = \bx).$$
		For a capture history $\by \neq \bzero$,
		\begin{align*}
			\Pb(\bY = \by | \bX = \bx) =& \  \frac{\Pb(\bY = \by| \bX = \bx)\Pb(\bY \neq \bzero| \bX = \bx)}{\Pb(\bY \neq \bzero| \bX = \bx)}\\
			=& \ \Pb(\bY = \by | \bX = \bx, \bY \neq \bzero)\gamma(\bx)\\
			=& \ \Qb(\bY = \by | \bX = \bx)\gamma(\bx).
		\end{align*}
		Substituting the above relation in the assumption statement and using the notation of the $q$-probabilities, we get 
		\begin{align*}
			or,\ & \Qb(Y_1 = 1,\, Y_2 = 1|\bX = \bx)\gamma(\bx) = \Qb(Y_1 = 1|\bX = \bx)\Qb(Y_2 = 1|\bX = \bx)\gamma(\bx)^2\\
			or,\ & \gamma(\bx) = \dfrac{q_{12}(\bx)}{q_{1}(\bx)q_{2}(\bx)}.
		\end{align*}
	\end{proof}
	
	\begin{proof}[\textbf{Proof of lemma \ref{lem:efficient influence function}}]
		
		There are several ways one can derive an efficient influence function and corresponding efficiency bound. Here, we first find a putative influence function in the discrete case using a standard Gateaux derivative argument. Then we show that this influence function is actually the efficient one in a general nonparametric model (continuous or discrete or mixed), by checking that the corresponding remainder term in a von Mises expansion is second-order.
		
		To find a candidate influence function, we consider a special parametric submodel (i.e., deviation from $\Qb$) given by $\Qb_\epsilon$ with density $q_\epsilon = (1 - \epsilon) q(\bz) + \epsilon \bar{q}(\bz)$ where $\bar{q}=\bar{q}(\bz) = \one(\bz=\tilde{\bz})$ is a point mass at $\bZ=\tilde{\bz}$, and for which the pathwise derivative
		$$ \frac{\partial}{\partial \epsilon}\left\{\frac{1}{\psi(\Qb_\epsilon)} \right\}\Bigg\vert_{\epsilon = 0} $$
		actually equals the influence function (in the discrete case) \citet{mises1947asymptotic, hampel1974influence}. We also let $q_{s,\epsilon}(\bx)$ denote the analog of $q_s(\bx)$ under the submodel for $s \in \{1,2,12\}$, e.g., the marginal density for $\bX$ under $\Qb_\epsilon$ is
		$$ q_\epsilon(\bx) = \sum_{\by} q_\epsilon(\bz) = (1-\epsilon) q(\bx) + \epsilon \one(\bx=\tilde{\bx}) . $$
		Now the above pathwise derivative equals
		\begin{align*}
			\frac{\partial}{\partial \epsilon}\left\{\frac{1}{\psi(\Qb_\epsilon)}\right\}\Bigg\vert_{\epsilon = 0} =& \ \frac{\partial}{\partial \epsilon} \int \frac{q_{2,\epsilon}(\bx) q_{1,\epsilon}(\bx)}{q_{12,\epsilon}(\bx)} q_\epsilon(\bx) d\bx \Bigg\vert_{\epsilon = 0} \\
			=& \ \int\frac{\partial}{\partial \epsilon}\left\{\dfrac{q_{2,\epsilon}(\bx) q_{1,\epsilon}(\bx)}{q_{12,\epsilon}(\bx)}q_\epsilon(\bx)\right\}d\bx\Bigg\vert_{\epsilon = 0} \\
			=& \ \int \dfrac{q_{2,\epsilon}(\bx) q_{1,\epsilon}(\bx)}{q_{12,\epsilon}(\bx)}q_\epsilon(\bx)\left\{ \frac{q_{1,\epsilon}'(\bx)}{q_{1,\epsilon}(\bx)} + \frac{q_{2,\epsilon}'(\bx)}{q_{2,\epsilon}(\bx)} - \frac{q_{12,\epsilon}'(\bx)}{q_{12,\epsilon}(\bx)} + \frac{q_{\epsilon}'(\bx)}{q_{\epsilon}(\bx)}\right\} d\bx \Bigg\vert_{\epsilon = 0}.
		\end{align*}
		where the last line follows by the product rule. For the discrete case, we use the notation of the integral to denote summation over $\bx$.
		
		For the derivatives appearing above, by the definition of $\bar{q}$, we have $q'_\epsilon(\bx) = \frac{\partial}{\partial \epsilon} q_\epsilon(\bx) = \one(\bx = \tilde\bx) - q(\bx)$. Similarly, by using derivative of product rule for $q'_{1,\epsilon}(\bx)$, we get
		\begin{align*}
			q'_{1,\epsilon}(\bx) =& \ \frac{\partial}{\partial\epsilon}q_{1,\epsilon}(\bx) = \frac{\partial}{\partial\epsilon}\frac{\Qb_\epsilon(Y_1 = 1, \bX = \bx)}{q_\epsilon(\bx)}\\
			=& \ \frac{\partial}{\partial\epsilon}\frac{(1 - \epsilon)\Qb(Y_1 = 1, \bX = \bx) + \epsilon\one(\tilde{Y}_1 = 1, \bx = \tilde\bx)}{(1 - \epsilon)q(\bx) + \epsilon\one(\bx = \tilde\bx)},\, \text{ where }\tilde{Y}_1 = \one(\tilde{y}_1 = 1)\\
			=& \ \frac{-\Qb(Y_1 = 1, \bX = \bx) + \one(\tilde{Y}_1 = 1, \bx = \tilde\bx)}{(1 - \epsilon)q(\bx) + \epsilon\one(\bx = \tilde\bx)}\\
			& \ - \frac{(1 - \epsilon)\Qb(Y_1 = 1, \bX = \bx) + \epsilon\one(\tilde{Y}_1 = 1, \bx = \tilde\bx)}{\{(1 - \epsilon)q(\bx) + \epsilon\one(\bx = \tilde\bx)\}^2}q'_\epsilon(\bx).
		\end{align*}
		The last step follows from the product rule of derivatives. Finally, setting $\epsilon = 0$, we get $q'_{1\epsilon}(\bx)|_{\epsilon = 0} = \ \frac{\one(\bx = \tilde\bx)}{q(\bx)}\{\tilde{Y}_1 - q_1(\bx)\}$. The derivatives for $q_{2,\epsilon}$ and $q_{12,\epsilon}$ follow similarly.
		
		Thus, combining the above results and using the discrete nature of the distribution, we get
		\begin{align*}
			\phi =& \ \frac{\partial}{\partial \epsilon}\left\{\frac{1}{\psi(\Qb_\epsilon)}\right\}\Bigg\vert_{\epsilon = 0}\\
			=& \ \sum_{\bx} \dfrac{q_1(\bx)q_2(\bx)}{q_{12}(\bx)}\one(\bx = \tilde\bx)\left\{\frac{\tilde{Y}_1 - q_{1}(\bx)}{q_1(\bx)} + \frac{\tilde{Y}_2 - q_{2}(\bx)}{q_2(\bx)} - \frac{\tilde{Y}_{1}\tilde{Y}_{2} - q_{12}(\bx)}{q_{12}(\bx)}\right\}\\
			& \ + \sum_\bx \dfrac{q_1(\bx)q_2(\bx)}{q_{12}(\bx)} \left\{\one(\bx = \tilde\bx) - q(\bx)\right\}\\
			= & \ \frac{1}{\gamma(\tilde\bx)} \left\{\dfrac{\tilde{Y}_1}{q_2(\tilde\bx)} + \dfrac{\tilde{Y}_2}{q_2(\tilde\bx)} - \dfrac{\tilde{Y}_1\tilde{Y}_2}{q_{12}(\tilde\bx)}\right\} - \frac{1}{\psi},\, \text{ using the definition of } \gamma(\bx).
		\end{align*}
		
		Now that we have a candidate influence function that is valid in a discrete model, we evaluate the remainder term $R_2$ in a general submodel, to show that it is actually the efficient influence function in the general case as well.
		
		Letting $\bar\psi = \psi(\bar\Qb)$ for a generic distribution $\bar\Qb$, the remainder of the so-called von Mises expansion \citep{bickel1993efficient, van2003unified} is then given by:
		\begin{align*}
			R_2(\Qb, \bar\Qb) \equiv& \ \frac{1}{\bar\psi} - \frac{1}{\psi} + \int \phi(\bz; \bar{\Qb})d\Qb \\
			=& \ \int \Bigg[\frac{1}{\bar\gamma(\bx)} \left\{\dfrac{Y_1}{\bar{q}_1(\bx)} + \dfrac{Y_2}{\bar{q}_2(\bx)} - \dfrac{Y_1Y_2}{\bar{q}_{12}(\bx)}\right\} - \frac{1}{\bar\psi} + \frac{1}{\bar\psi} - \frac{1}{\psi}\Bigg]d\Qb(\bx)\\
			=& \ \int \Bigg[\frac{1}{\bar\gamma(\bx)} \left\{\dfrac{q_1(\bx)}{\bar{q}_1(\bx)} + \dfrac{q_2(\bx)}{\bar{q}_2(\bx)} - \dfrac{q_{12}(\bx)}{\bar{q}_{12}(\bx)}\right\} - \frac{1}{\gamma(\bx)}\Bigg]d\Qb(\bx)\\
			=& \ \int \left\{\dfrac{q_1(\bx)\bar{q}_2(\bx)}{\bar{q}_{12}(\bx)} + \dfrac{\bar{q}_1(\bx)q_2(\bx)}{\bar{q}_{12}(\bx)} - \dfrac{q_{12}(\bx)}{\bar\gamma(\bx)\bar{q}_{12}(\bx)}- \frac{1}{\gamma(\bx)}\right\} \ d\Qb(\bx)\\
			=& \ \int \left\{\dfrac{q_1(\bx)\bar{q}_2(\bx)}{\bar{q}_{12}(\bx)} + \dfrac{\bar{q}_1(\bx)q_2(\bx)}{\bar{q}_{12}(\bx)} - \dfrac{\bar{q}_1(\bx)\bar{q}_2(\bx)}{\bar{q}_{12}(\bx)} - \dfrac{q_1(\bx)q_2(\bx)}{\bar{q}_{12}(\bx)}\right.\\
			& \ \left. + \  \dfrac{\bar{q}_1(\bx)\bar{q}_2(\bx)}{\bar{q}_{12}(\bx)} + \dfrac{q_1(\bx)q_2(\bx)}{\bar{q}_{12}(\bx)} - \dfrac{q_{12}(\bx)}{\bar\gamma(\bx)\bar{q}_{12}(\bx)}- \frac{1}{\gamma(\bx)}\right\} d\Qb(\bx)\\
			=& \ \int\frac{1}{\bar{q}_{12}(\bx)} \Bigg[\left\{q_1(\bx) - \bar{q}_1(\bx)\right\}\left\{\bar{q}_2(\bx) - q_2(\bx)\right\}\Bigg.\\
			& \ \ \Bigg. + \ \left\{q_{12}(\bx) - \bar{q}_{12}(\bx)\right\}\left\{\frac{1}{\gamma(\bx)} - \frac{1}{\bar{\gamma}(\bx)}\right\}\Bigg]d\Qb(\bx).
		\end{align*}
		
		The above shows that the remainder of the von Mises expansion is second-order, i.e., involving products of differences between components of $\Qb$ and $\bar\Qb$. This fact implies the more general pathwise differentiability condition that
		$$ \frac{\partial}{\partial \epsilon} \left\{\frac{1}{\psi(\Qb_\epsilon)}\right\}\Bigg|_{\epsilon = 0} = \int \phi\, \frac{\partial}{\partial \epsilon}\left\{log \ q_\epsilon(\bz)\Big\vert_{\epsilon = 0}\right\}d\Qb. $$
		for any smooth parametric submodel $\Qb_\epsilon$ \citep{kennedy2020efficient}. Thus, the candidate influence function satisfies the above pathwise differentiability condition and hence, is an efficient influence function. Moreover, since the model is non-parametric, $\phi$ is the only efficient influence function \citep{bickel1993efficient, tsiatis2006semiparametric, van2002semiparametric}. 
	\end{proof}
	
	\begin{proof}[\textbf{Proof of Theorem \ref{thm:var_phi}}]
		We will evaluate the variance of the efficient influence function $\phi \equiv \phi(\bZ; \Qb)$. Recall, $\bZ = (\bY, \bX)$. Hence, we use the law of total variance by conditioning on $\bX$. 
		\begin{align*}
			\var (\phi) =& \ \var \Bigg[\frac{1}{\gamma(\bX)}\left\{\dfrac{Y_1}{q_{1}(\bX)} +  \dfrac{Y_{2}}{q_{2}(\bX)} - \dfrac{Y_1Y_2}{q_{12}(\bX)}\right\} - \frac{1}{\psi}\Bigg]\\
			=& \ \var\Bigg(\E\Bigg[\frac{1}{\gamma(\bX)}\left\{\dfrac{Y_1}{q_{1}(\bX)} +  \dfrac{Y_{2}}{q_{2}(\bX)} - \dfrac{Y_1Y_2}{q_{12}(\bX)}\right\}\Bigg| \bX\Bigg]\Bigg)\\
			& \ + \E\Bigg(\var\Bigg[\frac{1}{\gamma(\bX)}\left\{\dfrac{Y_1}{q_{1}(\bX)} +  \dfrac{Y_{2}}{q_{2}(\bX)} - \dfrac{Y_1Y_2}{q_{12}(\bX)}\right\}\Bigg| \bX\Bigg]\Bigg), \, \text{by the law of total variance}\\
			=& \ \var\left\{\frac{1}{\gamma(\bX)}\right\} + \E\Bigg[\frac{1}{\gamma(\bX)^2}\Bigg\{\dfrac{1-q_1(\bX)}{q_{1}(\bX)} +  \dfrac{1 - q_2(\bX)}{q_{2}(\bX)} + \dfrac{1 - q_{12}(\bX)}{q_{12}(\bX)}\\
			& \ + 2\ \dfrac{q_{12}(\bX) - q_1(\bX)q_2(\bX)}{q_{1}(\bX)q_2(\bX)} - 2 \ \dfrac{q_{12}(\bX) - q_{1}(\bX)q_{12}(\bX)}{q_{1}(\bX)q_{12}(\bX)} - 2 \ \dfrac{q_{12}(\bX) - q_{2}(\bX)q_{12}(\bX)}{q_2(\bX)q_{12}(\bX)}\Bigg\}\Bigg]\\
			=& \ \var\left\{\frac{1}{\gamma(\bX)}\right\} + \E\Bigg[\frac{1}{\gamma(\bX)^2}\Bigg\{2\gamma(\bX) - \dfrac{1}{q_{1}(\bX)} -  \dfrac{1}{q_{2}(\bX)} + \dfrac{1}{q_{12}(\bX)} - 1\Bigg\}\Bigg].
		\end{align*}
		The last two equalities follow by evaluating the conditional expectation and the variance the two terms respectively and the relation that $\gamma(\bx) = \frac{q_{12}(\bx)}{q_1(\bx)q_2(\bx)}$. By simple algebra, we can further simplify the variance as follows
		\begin{align*}
			\var (\phi) =& \ \var\left\{\frac{1}{\gamma(\bX)}\right\} + \E\Bigg[\frac{1}{\gamma(\bX)^2}\Bigg\{2\gamma(\bX)-\dfrac{1}{q_{1}(\bX)} -  \dfrac{1}{q_{2}(\bX)} + \dfrac{1}{q_{12}(\bX)} - 1\Bigg\}\Bigg]\\
			=& \ \var\left\{\frac{1}{\gamma(\bX)}\right\} + \E\Bigg[\frac{1}{\gamma(\bX)^2}\Bigg\{2\gamma(\bX) - \dfrac{1 + q_{12}(\bX) - q_0(\bX)}{q_{1}(\bX)q_{2}(\bX)} + \dfrac{1}{q_{12}(\bX)} - 1\Bigg\}\Bigg]\\
			=& \ \var\left\{\frac{1}{\gamma(\bX)}\right\} + \E\Bigg(\frac{1}{\gamma(\bX)}\Bigg[\left\{\frac{1}{\gamma(\bX)} - 1\right\}\Bigg\{\dfrac{1}{q_{12}(\bX)} - 1\Bigg\} + \dfrac{q_0(\bX)}{q_{12}(\bX)}\Bigg]\Bigg).
		\end{align*}
		\textit{Recall: $q_0(\bX) = \Qb(Y_1 = 0, Y_2 = 0\mid \bY \neq \bzero, \bX = \bX)$.}
	\end{proof}
	
	\begin{proof}[\textbf{Proof of Theorem \ref{thm:bc_errorterm}}]
		Let $\Em_N = \widehat\psi^{-1}_{dr} - \psi^{-1} - \Qn\phi$. We can expand it as follows.
		\begin{align*}
			\Em_N =& \ \frac{1}{\widehat\psi_{dr}} - \frac{1}{\psi} - \Qn\phi\\
			=& \ \frac{1}{\widehat\psi_{pi}} + \Qb_N\widehat\phi - \frac{1}{\psi} - \Qn\phi,\quad \text{using the definition in equation \ref{eq:biascorrection}}\\
			=& \ -\int\widehat\phi(\bz)d\Qb(\bz) + \widehat{R}_2 + \Qn\widehat\phi - \Qn\phi,\quad \text{using von Mises expansion of $\widehat\psi^{-1}$}.
		\end{align*}
		The last step follows using $\frac{1}{\widehat\psi} - \frac{1}{\psi} = -\int\widehat\phi(\bz) d\Qb(\bz) + \widehat{R}_2$ from the proof of lemma \ref{lem:efficient influence function}. The formula of $\widehat{R}_2$ is presented in the proof of lemma \ref{lem:efficient influence function}.
		
		For simplicity of notation in the proof we will use $\Qb(\cdot) = \E(\cdot | \bY \neq\bzero, \bZ^n) = \int (\cdot) d\Qb(\bz)$ to denote the observed population average function conditioned on the training sample $\bZ^n$. Using that fact that $\Qb\phi = 0$, we get
		\begin{align*}
			\Em_N  =& \ \widehat{R}_2 + (\Qb_N - \Qb)(\widehat\phi - \phi).
		\end{align*}
		We want to bound $\Em_N$ by presenting a bound on $\E(\Em_N^2)$. We will show that the expected value of $\Em$ is expected value of the error term $\widehat{R}_2$.
		\begin{align*}\E(\Em_N) =& \  \E\left\{\E\left(\Em_N| \bZ^n\right)\right\}\\
			=& \ \E\left[\E\left\{\widehat{R}_2 + (\Qb_N - \Qb)(\widehat\phi - \phi)\big| \bZ^n \right\}\right]\\
			=& \ \E(\widehat{R}_2) + \E\left[\E\left\{(\Qb_N - \Qb)(\widehat\phi - \phi)\big| \bZ^n \right\}\right].
		\end{align*}
		Next, we will show that the second quantity is 0.
		\begin{align*}
			\E\left[\E\left\{(\Qb_N - \Qb)(\widehat\phi - \phi)\big| \bZ^n \right\}\right] =& \ \E\left[\E\left\{\Qb_N(\widehat\phi-\phi) - \Qb(\widehat\phi-\phi)\big| \bZ^n \right\}\right]\\
			=& \ \E\left\{\E\left(\widehat\phi - \phi\big| \bZ^n\right) - \E\left(\widehat\phi - \phi\big| \bZ^n \right)\right\}\\
			=& \ 0.
		\end{align*}
		The second equality follows because $\Qb_N$ is evaluated on i.i.d. terms and $\Qb (\widehat\phi - \phi) = \E(\widehat\phi-\phi |\bZ^n)$ by definition. Next, we evaluate the variance of $\Em_N$.
		\begin{align*}
			\var(\Em_N) =& \ \E\{\var(\Em_N\vert\bZ^n)\} + \var\{\E(\Em_N\vert\bZ^n)\},\,\text{ by the law of total variance}\\
			=& \ \E\{\var(\Em_N\vert\bZ^n)\} + \var(\widehat{R}_2)\\
			=& \ \E\left[\var\left\{\widehat{R}_2 + (\Qb_N - \Qb)(\widehat\phi-\phi)\Big\vert\bZ^n\right\}\right] + \var(\widehat{R}_2)\\
			=& \ \E\left[\var\left\{\Qb_N(\widehat\phi-\phi)\Big\vert\bZ^n\right\}\right] + \var(\widehat{R}_2).
		\end{align*}
		The last equality follows because $\widehat{R}_2$ and $\Qb(\cdot)$ are constants when conditioned on $\bZ^n$. Now,
		\begin{align*}
			\var\left\{\Qb_N(\widehat\phi - \phi)\big|\bZ^n\right\} =& \ \frac{1}{N} \var(\widehat\phi - \phi |\bZ^n),\, \text{property of variance of sample average}\\
			\le& \ \frac{1}{N} \Vert\widehat\phi - \phi\Vert^2.
		\end{align*}
		Thus, combining the above results we get
		$\E(\Em_N^2) \le \E(\widehat{R}_2^2) + \frac{\E\Vert\widehat\phi-\phi\Vert^2}{N}$.
		
		Hence, by using Markov's inequality combined with the above inequality gives the required result.
		$$\Pb\left(\left|\widehat\psi^{-1}_{dr} - \psi^{-1} - \Qn\phi\right|\leq\delta \right) \geq 1 - \frac{1}{\delta^2}\E(\Em_N^2).$$
	\end{proof}
	
	\begin{proof}[\textbf{Proof of Theorem \ref{thm:normality}}]
		The error in the estimation of $\psi^{-1}$ is $\widehat\psi^{-1} - \psi^{-1}$. Following the proof of Theorem \ref{thm:bc_errorterm}, it can be expressed as 
		$$\widehat\psi^{-1} - \psi^{-1} = (\Qn - \Qb)\widehat{\phi} + \widehat{R}_2,$$
		where $\Qb\widehat\phi$ denotes $\int\widehat\phi(\bz)d\Qb(\bz)$ as mentioned in the proof of Theorem \ref{thm:bc_errorterm}.
		
		Let $\widetilde\sigma^2 = \var(\widehat\phi \mid \bZ^n)$, and note $\widehat\psi^{-1} - \psi^{-1} - \widehat{R}_2 = (\Qn-\Qb)\widehat\phi$ is a sample average of a fixed function given the training sample $\bZ^n$. Therefore by Berry-Esseen we have for any $t'$ and $N$ that 
		\begin{align*}
			\Phi(t')-\frac{C \rho}{\widetilde\sigma^3 \sqrt{N}} &\leq \Pb\left( \frac{\widehat\psi^{-1} - \psi^{-1} - \widehat{R}_2}{\widetilde\sigma / \sqrt{N}} \leq t' \Bigm| \bZ^n \right) \leq \Phi(t') +  \frac{C \rho}{\widetilde\sigma^3 \sqrt{N}}
		\end{align*}
		Taking $t' = \frac{\widehat\sigma}{\widetilde\sigma}  t - \frac{\widehat{R}_2}{\widetilde\sigma/\sqrt{N}} $ for $\widehat\sigma^2 = \widehat{\var}(\widehat{\phi})$ and $\rho = \E\left(|\widehat{\phi} - \Qb\widehat{\phi}|^3\Big\vert \bZ^n\right)$, this implies
		\begin{align*}
			\Phi\left( \frac{\widehat\sigma}{\widetilde\sigma}  t - \frac{\widehat{R}_2}{\widetilde\sigma/\sqrt{N}}  \right)-\frac{C \rho}{\widetilde\sigma^3 \sqrt{N}} &\leq \Pb\left( \frac{\widehat\psi^{-1} - \psi^{-1} }{\widehat\sigma / \sqrt{N}} \leq t  \Bigm| \bZ^n \right) \leq \Phi\left( \frac{\widehat\sigma}{\widetilde\sigma}  t - \frac{\widehat{R}_2}{\widetilde\sigma/\sqrt{N}} \right) +  \frac{C \rho}{\widetilde\sigma^3 \sqrt{N}}
		\end{align*}
		Now note  by the mean value theorem, for some $t_n $ between $t$ and $t+\left( \frac{\widehat\sigma}{\widetilde\sigma} - 1 \right) t - \frac{\widehat{R}_2}{\widetilde\sigma/\sqrt{N}} $, we have that
		\begin{align*}
			\left\vert\Phi\left( t + \left( \frac{\widehat\sigma}{\widetilde\sigma} - 1 \right) t - \frac{\widehat{R}_2}{\widetilde\sigma/\sqrt{N}} \right) -  \Phi(t)\right\vert &= \left\vert\Phi'(t_n) \left( \left( \frac{\widehat\sigma}{\widetilde\sigma} - 1 \right) t - \frac{\widehat{R}_2}{\widetilde\sigma/\sqrt{N}} \right)\right\vert \\
			&\leq \frac{1}{\sqrt{2\pi}} \left( \left| \frac{\widehat\sigma}{\widetilde\sigma} - 1  \right| |t| + \frac{|\widehat{R}_2|}{\widetilde\sigma/\sqrt{N}} \right)\equiv \Delta_n
		\end{align*}
		where the second inequality used the fact that $\sup_t \Phi'(t) \leq 1/\sqrt{2\pi}$ and the triangle inequality.
		
		Therefore
		\begin{align*}
			- \Delta_n -\frac{C \rho}{\widetilde\sigma^3 \sqrt{N}} &\leq \Pb\left( \frac{\widehat\psi^{-1} - \psi^{-1} }{\widehat\sigma / \sqrt{N}}  \leq t  \Bigm| \bZ^n \right) - \Phi(t) \leq  \Delta_n +  \frac{C \rho}{\widetilde\sigma^3 \sqrt{N}}
		\end{align*}
		
		This implies by iterated expectation that 
		\begin{align*}
			\left| \Pb\left( \frac{\widehat\psi^{-1} - \psi^{-1} }{\widehat\sigma / \sqrt{N}} \leq t   \right) - \Phi(t) \right| &\leq  \sqrt{\frac{1}{2\pi}} \left\{ \sqrt{N}\E\left(  \frac{ |\widehat{R}_2|}{ \widetilde\sigma} \right) + |t| \E\left(  \left| \frac{\widehat\sigma}{\widetilde\sigma} - 1  \right| \right) \right\} + \frac{C}{\sqrt{N}} \E\left( \frac{\rho}{\widetilde\sigma^3 }  \right) .
		\end{align*}
		
		\bigskip
		
		Therefore if $\widetilde\sigma \gtrsim 1$ with probability one, $\E(\rho) <c'$ for some $c'>0$, $\E|\widetilde\sigma - \widehat\sigma| \lesssim n^{-1/2}$ and $\E|\widehat{R}_2| \lesssim n^{-2\beta}$, then
		$$ \left| \Pb\left( \frac{\widehat\psi^{-1} - \psi^{-1} }{\widehat\sigma / \sqrt{N}} \leq t   \right) - \Phi(t) \right| \lesssim n^{-1/2} + n^{(1-4\beta)/2}  $$
		If $\beta > 1/4$, then there exists an $N=N_\epsilon$ guaranteeing that the LHS is no more than $\epsilon$.\\[2ex]
		\textbf{Bound on $\mathbf{\E|\widehat\sigma - \widetilde\sigma|}$}\\[2ex]
		We will show that this bound is $n^{-1/2}$. For $\widehat\phi$, we have defined the quantities $\widetilde\sigma^2 = \var(\widehat\phi \mid \bZ^n)$ and $\widehat\sigma^2 = \widehat{\var}(\widehat{\phi})$. We can further expand $\widetilde\sigma^2$ as follows
		\begin{align*}
			\widetilde\sigma^2 = \var(\widehat\phi \mid \bZ^n) = \E\left(\widehat\phi^2 | \bZ^n\right) - \left\{\E\left(\widehat\phi|\bZ^n\right)\right\}^2.
		\end{align*}
		The second quantity $\widehat\sigma^2$ is the unbiased estimator of the variance of $\widehat\phi$ given the test sample i.e. $\frac{N}{N-1} \left\{\Qn\widehat{\phi}^2 - (\Qn\widehat{\phi})^2\right\}$. Then by iterated expectation
		\begin{align*}
			\E(\widehat\sigma^2) = \E\{ \E(\widehat\sigma^2 \mid \bZ^n) \} = \E(\widetilde\sigma^2),
		\end{align*}
		where the second equality follows from the unbiasedness property.
		
		We need the bound on $\left|\widehat\sigma - \widetilde \sigma\right|$ which can be expressed as a linear function of the absolute difference of the respective squares.
		$$|\widehat\sigma - \widetilde\sigma| = |\widehat\sigma - \widetilde\sigma|\frac{|\widehat\sigma + \widetilde\sigma|}{|\widehat\sigma + \widetilde\sigma|} \leq \epsilon^{-1}|\widehat\sigma^2 - \widetilde\sigma^2|,\, \text{ if } \widehat\sigma + \widetilde\sigma > \epsilon >0.$$
		Hence, it is enough to show the bound on $|\widehat\sigma^2 - \widetilde\sigma^2|$. Equivalently, we can show that $\E|\widehat\sigma^2 - \widetilde\sigma^2|^2 \lesssim n^{-1}$. Evaluating this quantity by the law of iterated expectation
		\begin{align*}
			\E\left(\widehat\sigma^2 - \widetilde \sigma^2\right)^2 =& \ \E\left[\E\left\{\left(\widehat\sigma^2 - \widetilde \sigma^2\right)^2\big\vert \bZ^n\right\}\right]\\
			=& \ \E\left\{\E\left(\widehat\sigma^4 + \widetilde \sigma^4 - 2\widehat\sigma^2\widetilde \sigma^2\big\vert \bZ^n\right)\right\}\\
			=& \ \E\left\{\E\left(\widehat\sigma^4\big\vert \bZ^n\right) - \widetilde \sigma^4\right\},
		\end{align*}
		where the last equality follows from $\E(\widehat\sigma^2|\bZ^n) = \widetilde\sigma^2$. Next, we $\widehat\sigma^4$.
		\begin{align*}
			\widehat\sigma^4 =& \  \left(\widehat\sigma^2\right)^2\\
			=& \  \frac{N^2}{(N-1)^2}\left\{\frac{\sum_i \widehat{\phi}^2}{N} - \left(\frac{\sum_i \widehat{\phi}}{N}\right)^2\right\}^2\\
			=& \  \frac{N^2}{(N-1)^2}\left\{\frac{\sum_i \widehat{\phi}_i^4 + \sum_{i\neq j} \widehat{\phi}_i^2 \widehat{\phi}_j^2}{N^2} - 2\frac{\sum_i \widehat{\phi}_i^4 + \sum_{i\neq j}\widehat{\phi}_i^2\widehat{\phi}_j^2 + 2\sum_{i\neq j} \widehat{\phi}_i^3\widehat{\phi}_j + \sum_{i\neq j\neq k}\widehat{\phi}_i^2\widehat{\phi}_j\widehat{\phi}_k}{N^3}\right.\\
			& \ + \left. \frac{\sum_i \widehat{\phi}_i^4 + 3\sum_{i\neq j}\widehat{\phi}_i^2\widehat{\phi}_j^2 + 4 \sum_{i\neq j} \widehat{\phi}_i^3\widehat{\phi}_j + 6\sum_{i\neq j\neq k} \widehat{\phi}_i^2\widehat{\phi}_j\widehat{\phi}_k + \sum_{i\neq j\neq k\neq l}\widehat{\phi}_i\widehat{\phi}_j\widehat{\phi}_k\widehat{\phi}_l}{N^4}\right\}.
		\end{align*}
		Let $\mu_m = \E(\widehat{\phi}^m\mid \bZ^n)$. Thus, $\E(\widehat\sigma^2\mid \bZ^n) = \widetilde \sigma^2 = \mu_2 - \mu_1^2$.
		\begin{align*}
			\E\left(\widehat\sigma^4\mid \bZ^n\right) =& \  \frac{N^2}{(N-1)^2}\left\{\frac{N\mu_4 + N(N-1)\mu_2^2}{N^2}\right.\\
			& \  - 2\frac{N\mu_4 + N(N-1)\mu_2^2 + 2N(N-1)\mu_3\mu_1 + N(N-1)(N-2)\mu_2\mu_1^2}{N^3}\\
			& \ + \left. \frac{\mu_4 + 3(N-1)\mu_2^2 + 4(N-1)\mu_3\mu_1 + 6(N-1)(N-2)\mu_2\mu_1^2 + (N-1)(N-2)(N-3)\mu_1^4}{N^3}\right\}\\
			=& \  \frac{\mu_4}{(N-1)^2}\left\{N - 2 + \frac{1}{N}\right\} + \frac{\mu_2^2}{N-1}\left\{N - 2 + \frac{3}{N}\right\} +\frac{\mu_3\mu_1}{N-1}\left\{-2 + \frac{4}{N}\right\}\\
			& \ + \frac{\mu_2\mu_1^2}{N-1}\left\{-2(N-2) + \frac{6(N-2)}{N}\right\} + \frac{\mu_1^4}{N-1}\frac{(N-2)(N-3)}{N}\\
			=& \ \frac{\mu_4}{N} + \frac{\mu_2^2(N^2 - 2N + 3)}{N(N-1)} - 2\frac{\mu_3\mu_1(N-2)}{N(N-1)} - \frac{2\mu_2\mu_1^2(N-2)(N-3)}{N(N-1)} + \frac{\mu_1^4(N-2)(N-3)}{N(N-1)}.
		\end{align*}
		
		Thus, combining the results
		\begin{align*}
			& \ \E\left(\widehat\sigma^4\big\vert \bZ^n\right) - \widetilde\sigma^4\\
			=& \  \frac{\mu_4}{N} + \frac{\mu_2^2(N^2 - 2N + 3)}{N(N-1)} - 2\frac{\mu_3\mu_1(N-2)}{N(N-1)} - \frac{2\mu_2\mu_1^2(N-2)(N-3)}{N(N-1)} + \frac{\mu_1^4(N-2)(N-3)}{N(N-1)}\\
			& \ - (\mu_2 - \mu_1^2)^2,\,\,\text{ since }\E(\widehat\sigma^2\mid \bZ^n) = \widetilde \sigma^2 = \mu_2 - \mu_1^2\\
			=& \ \frac{\mu_4}{N} - \frac{\mu_2^2}{N(N-1)}(N - 3) + \frac{4\mu_2\mu_1^2}{N(N-1)}(2N - 3) - \frac{2\mu_1^4}{N(N-1)}(2N - 3) - 2\frac{\mu_3\mu_1}{N(N-1)}(N-2)\\
			\lesssim& \ N^{-1} \lesssim n^{-1}.
		\end{align*}
		Thus, $\left\vert \widehat\sigma^2 -\widetilde\sigma^2\right\vert \lesssim \frac{1}{\sqrt{n}}$.
	\end{proof}
	
	\begin{proof}[\textbf{Proof of Theorem \ref{thm:confidenceinterval}}] The estimate of population size, $\widehat{n} = N/\widehat\psi$ depends on two random quantities (i) the number of observations $N$ and (ii) the estimate of the capture probability $\psi$.\\[2ex]
		\textbf{Calculation of mean and variance of $\widehat{n}$}\\
		First, we re-write $\frac{N}{\widehat\psi}$ as a sample average for ease of calculation. As mentioned in the proof of Theorem \ref{thm:bc_errorterm}, we use $\Qb\widehat\varphi$ to denote the population average conditioned on training sample $\bZ^n$ i.e., $\int\widehat\varphi(\bz)d\Qb(\bz)$.
		\begin{align*}
			\widehat{n} - n &= \ N/\widehat\psi - N/\psi + N/\psi - n\\
			&= \ N\left\{(\Qn - \Qb)\widehat{\varphi} + \widehat{R}_2\right\} + N/\psi - n, \ \text{(follows from proof of Theorem \ref{thm:bc_errorterm})}\\
			&= \ n\Pn\one(\bY \neq \bzero)\left(\widehat\varphi - \Qb\widehat\varphi + \widehat{R}_2 + \psi^{-1}\right) - n, \ \text{since } N = n\Pn\one(\bY \neq \bzero)\\
			&= \ n\Pn \underbrace{\left[\one(\bY \neq \bzero)\left(\widehat{\varphi} - \Qb\widehat{\varphi}\right) + \left\{\one(\bY \neq \bzero) - \psi\right\}\left(\widehat{R}_2 + \psi^{-1}\right)\right]}_{\zeta} + n\psi\widehat{R}_2.
		\end{align*}
		Thus, $\widehat{n} - n - n\psi\widehat{R}_2$ is a sample average $n\Pn\zeta$. Moreover, when we condition on the training sample $\bZ^n$, the $\zeta$'s are i.i.d. We present the conditional mean and variance of $\zeta$ below.
		\begin{align*}
			\E(\zeta | \bZ^n) &= \E\{\E(\zeta |\bY \neq \bzero, \bZ^n) \vert \bZ^n\} = 0.\\[1.5ex]
			\var(\zeta | \bZ^n) &= \var\{\E(\zeta | \bY \neq \bzero, \bZ^n)\} + \E\{\var(\zeta | \bY \neq \bzero, \bZ^n)\}\\
			&= \var\left[\left\{\one(\bY \neq \bzero) - \psi\right\} (\widehat{R}_2 + \psi^{-1}) \big\vert\bZ^n\right] + \E\left[ \one(\bY \neq \bzero) \var\left(\widehat{\varphi}| \bZ^n\right) \big\vert \bZ^n\right]\\
			&= \psi(1 - \psi)(\widehat{R}_2 + \psi^{-1})^2 + \psi \ \var\left(\widehat{\varphi}| \bZ^n\right)\\
			&= \frac{1 - \psi}{\psi}(\psi\widehat{R}_2 + 1)^2 + \psi \widetilde\varsigma^2,\,\, \text{ defining } \widetilde\varsigma^2 = \var(\widehat{\varphi}|\bZ^n).
		\end{align*}
		The population expectation of $\widehat{n} - n$ is $n\psi \E(\widehat{R}_2).$ And the population variance is presented below.
		\begin{align*}
			\var(\widehat{n} - n) &= \var\{\E(\widehat{n} - n | \bZ^n)\} + \E\{\var(\widehat{n} - n | \bZ^n)\}\\
			&= \var\{\E(n\Pn \zeta + n\psi \widehat{R}_2 | \bZ^n)\} + \E\{\var(n\Pn \zeta + n\psi \widehat{R}_2 | \bZ^n)\}\\
			&= \var(n\psi \widehat{R}_2) + \E\{n \ \var(\zeta | \bZ^n)\}\\
			&= n^2\psi^2 \var(\widehat{R}_2) + n\psi \E\left(\widetilde\varsigma^2 \right) + n\frac{1 - \psi}{\psi}\E(\psi\widehat{R}_2 + 1)^2.
		\end{align*}
		Thus, $\E(\widehat{n} - n)^2 = n^2\psi^2 \E(\widehat{R}_2^2) + n\psi \E\left(\widetilde\varsigma^2 \right) + n\frac{1 - \psi}{\psi}\E(\psi\widehat{R}_2 + 1)^2$.\\
		If $\E|\widehat{R}_2|$ is sufficiently small, then $\widehat{n} - n$ has expectation approximately 0 and variance approximately $n\psi\varsigma^2 + n(1 - \psi)/\psi$, where $\varsigma^2 = \var(\varphi)$.\\\\
		\textbf{Approximate normality}\\
		
		We define the estimated variance of $\widehat{n} - n - n\psi\widehat{R}_2$ conditioned on the training sample as $\widehat{n}\widehat\tau^2$, where $\widehat\tau^2 = \widehat\psi\widehat\varsigma^2 + \frac{1 - \widehat\psi}{\widehat\psi}$ and $\widehat\varsigma^2 = \widehat\var(\widehat\varphi | \bZ^n)$ is the unbiased estimator of $\widetilde\varsigma^2$ conditioned on the training data. Let $\widetilde\tau^2 = \var(\zeta | \bZ^n) = \psi \widetilde\varsigma^2 + \frac{1 - \psi}{\psi}(\psi\widehat{R}_2 + 1)^2$.\\\\
		
		We ultimately want to see the error in normal approximation for $\frac{\widehat{n} - n}{\widehat\tau\sqrt{\widehat{n}}}$.
		
		By Berry-Esseen we have for any $t'$ and $n$ that 
		\begin{align*}
			\Phi(t')-\frac{C \rho}{\widetilde\tau^3 \sqrt{n}} &\leq \Pb\left( \frac{\widehat{n} - n - n\psi\widehat{R}_2}{\widetilde\tau \sqrt{n}} \leq t' \Bigm| \bZ^n \right) \leq \Phi(t') +  \frac{C \rho}{\widetilde\tau^3 \sqrt{n}}
		\end{align*}
		Taking $t' = \frac{\widehat\tau\sqrt{\widehat{n}}}{\widetilde\tau\sqrt{n}} t - \frac{\psi\widehat{R}_2}{\widetilde\tau/\sqrt{n}} $ and $\rho = \E\left(|\zeta|^3\big\vert \bZ^n\right)$, this implies
		\begin{align*}
			\Phi\left(\frac{\widehat\tau\sqrt{\widehat{n}}}{\widetilde\tau\sqrt{n}} t - \frac{\psi\widehat{R}_2}{\widetilde\tau/\sqrt{n}}  \right)-\frac{C \rho}{\widetilde\tau^3 \sqrt{n}} &\leq \Pb\left( \frac{\widehat{n} - n }{\widehat\tau \sqrt{\widehat{n}}} \leq t  \Bigm| \bZ^n \right) \leq \Phi\left(\frac{\widehat\tau\sqrt{\widehat{n}}}{\widetilde\tau\sqrt{n}} t - \frac{\psi\widehat{R}_2}{\widetilde\tau/\sqrt{n}} \right) +  \frac{C \rho}{\widetilde\tau^3 \sqrt{n}}
		\end{align*}
		or equivalently
		\begin{align*}
			\Phi\left( t\frac{\widehat\tau\sqrt{\widehat{n}}}{\widetilde\tau\sqrt{n}} - \frac{\psi\widehat{R}_2}{\widetilde\tau/\sqrt{n}} \right) - \Phi(t) -\frac{C \rho}{\widetilde\tau^3 \sqrt{n}} \leq& \ \Pb\left( \frac{\widehat{n} - n }{\widehat\tau \sqrt{\widehat{n}}} \leq t  \Bigm| \bZ^n \right) - \Phi(t)\\
			\leq& \ \Phi\left( t\frac{\widehat\tau\sqrt{\widehat{n}}}{\widetilde\tau\sqrt{n}} - \frac{\psi\widehat{R}_2}{\widetilde\tau/\sqrt{n}}\right) - \Phi(t) +  \frac{C \rho}{\widetilde\tau^3 \sqrt{n}}.
		\end{align*}
		Now note by the mean value theorem, for some $t_n $ between $t$ and $t\frac{\widehat\tau\sqrt{\widehat{n}}}{\widetilde\tau\sqrt{n}} - \frac{\psi\widehat{R}_2}{\widetilde\tau/\sqrt{n}} $, we have that
		\begin{align*}
			\left|\Phi\left( t\frac{\widehat\tau\sqrt{\widehat{n}}}{\widetilde\tau\sqrt{n}} - \frac{\psi\widehat{R}_2}{\widetilde\tau/\sqrt{n}} \right) -  \Phi(t)\right| &= \left|\Phi'(t_n) \left( \left(\frac{\widehat\tau\sqrt{\widehat{n}}}{\widetilde\tau\sqrt{n}}- 1 \right) t - \frac{\psi\widehat{R}_2}{\widetilde\tau/\sqrt{n}} \right)\right| \\
			&\leq \frac{1}{\sqrt{2\pi}} \left( \left|\frac{\widehat\tau\sqrt{\widehat{n}}}{\widetilde\tau\sqrt{n}}- 1  \right| |t| + \frac{\psi|\widehat{R}_2|}{\widetilde\tau/\sqrt{n}} \right)\equiv \Delta_n
		\end{align*}
		where the second inequality used the fact that $\sup_t \Phi'(t) \leq 1/\sqrt{2\pi}$ and the triangle inequality.\\[2ex]
		
		Therefore
		\begin{align*}
			- \Delta_n -\frac{C \rho}{\widetilde\tau^3 \sqrt{n}} &\leq \Pb\left( \frac{\widehat{n} - n }{\widehat\tau \sqrt{\widehat{n}}}  \leq t  \Bigm| \bZ^n \right) - \Phi(t) \leq  \Delta_n +  \frac{C \rho}{\widetilde\tau^3 \sqrt{n}}
		\end{align*}
		This implies by iterated expectation that 
		\begin{align*}
			\left| \Pb\left( \frac{\widehat{n} - n }{\widehat\tau \sqrt{\widehat{n}}} \leq t   \right) - \Phi(t) \right| &\leq  \sqrt{\frac{1}{2\pi}} \left\{ \frac{\sqrt{n} \psi\E|\widehat{R}_2|}{ \widetilde\tau} + |t| \E\left(  \left|\frac{\widehat\tau\sqrt{\widehat{n}}}{\widetilde\tau\sqrt{n}}- 1  \right| \right) \right\} + \frac{C}{\sqrt{n}} \E\left( \frac{\rho}{\widetilde\tau^3} \right).
		\end{align*}\\\\
		\textbf{Bound on $\mathbf{\E\left|\frac{\widehat\tau\sqrt{\widehat{n}}}{\widetilde\tau\sqrt{n}} - 1\right|}$}\\[1.5ex]
		It is easy to see that $$\frac{1}{\sqrt{n}\widetilde\tau}\left|\widehat\tau\sqrt{\widehat{n}} - \widetilde\tau\sqrt{n}\right| = \frac{1}{\sqrt{n}\widetilde\tau}\left|\widehat\tau\sqrt{\widehat{n}} - \widetilde\tau\sqrt{n}\right|\frac{\left|\widehat\tau\sqrt{\widehat{n}} + \widetilde\tau\sqrt{n}\right|}{\left|\widehat\tau\sqrt{\widehat{n}} + \widetilde\tau\sqrt{n}\right|} \leq \frac{\left|\widehat\tau^2\widehat{n} - \widetilde\tau^2n\right|}{n\widetilde\tau^2}.$$
		By simple algebra, we can bound the quantity on the right hand side above as follows.
		\begin{align*}
			\frac{1}{n\widetilde\tau^2} \left|\widehat\tau^2\frac{N}{\widehat\psi} - \widetilde\tau^2n\right| =& \  \frac{1}{n\widetilde\tau^2} \left|\widehat\psi \widehat\varsigma^2\frac{N}{\widehat\psi} + \frac{1 - \widehat\psi}{\widehat\psi}\frac{N}{\widehat\psi} - \psi\widetilde\varsigma^2n - \frac{1 - \psi}{\psi}(\psi\widehat{R}_2 + 1)^2n\right|\\
			\leq& \ \frac{N}{n\widetilde\tau^2} |\widehat\varsigma^2 - \widetilde\varsigma^2| + \frac{1}{n\psi}|N - n\psi| + \frac{N}{n\widetilde\tau^2}\left|\frac{1}{\widehat\psi^2} - \frac{1}{\psi^2} - \frac{1}{\widehat\psi} + \frac{1}{\psi} \right| + \frac{1}{\widetilde\tau^2}(1 - \psi)\left(\psi\widehat{R}_2^2 + 2|\widehat{R}_2|\right).
		\end{align*}
		The last inequality follows using the definition of $\widetilde\tau$ and triangle inequality. To evaluate the third term, we will use the relation $\widehat\psi^{-1} - \psi^{-1} = (\Qn - \Qb)\widehat\varphi + \widehat{R}_2$. Thus,
		\begin{align*}
			& \ \frac{1}{\widehat\psi^2} - \frac{1}{\psi^2} - \frac{1}{\widehat\psi} + \frac{1}{\psi}\\
			=& \ \left\{\psi^{-1} + (\Qn - \Qb)\widehat\varphi + \widehat{R}_2\right\}^2 - \psi^{-2} - (\Qb - \Qn)\widehat\varphi - \widehat{R}_2\\
			=& \ \{(\Qn - \Qb)\widehat\varphi\}^2 + \widehat{R}_2^2 + (2\psi^{-1} + 2\widehat{R}_2 - 1)(\Qn - \Qb)\widehat\varphi + (2\psi^{-1} - 1)\widehat{R}_2.
		\end{align*}
		Similar to the proof of Theorem \ref{thm:bc_errorterm}, $$\E\left\{\big|(\Qn - \Qb)\widehat\varphi\big| \bigg|\bZ^n\right\} \leq \left(\E\left[\big\{\left(\Qn - \Qb\right)\widehat\varphi\big\}^2 \bigg| \bZ^n\right]\right)^{1/2} = \frac{\widetilde\varsigma}{\sqrt{N}} \leq \frac{\widetilde\tau}{\sqrt{\psi N}}.$$   
		The last bound follows by the relation between $\widetilde\tau$ and $\widetilde\varsigma$. Thus,
		\begin{align*}
			& \ \E\left\{\left|\frac{1}{\widehat\psi^2} - \frac{1}{\psi^2} - \frac{1}{\widehat\psi} + \frac{1}{\psi}\right|\Bigg|\bZ^n\right\}\\
			\leq& \ \E\left[\{(\Qn - \Qb)\widehat\varphi\}^2\big|\bZ^n\right] + \widehat{R}_2^2 + \left|2\psi^{-1} + 2\widehat{R}_2 - 1\right|\E\left\{\big|(\Qn - \Qb)\widehat\varphi\big|\bigg|\bZ^n\right\} + (2\psi^{-1} - 1)\left|\widehat{R}_2\right|\\
			\leq& \ \frac{\widetilde\tau^2}{N\psi} + \widehat{R}_2^2 + \left|2\psi^{-1} + 2\widehat{R}_2 - 1\right|\frac{\widetilde\tau}{\sqrt{N\psi}} + (2\psi^{-1} - 1)\left|\widehat{R}_2\right|.
		\end{align*}
		Combining the results above, we get the following bound.
		\begin{align*}
			\E\left|\frac{\widehat\tau \sqrt{\widehat{n}}}{\widetilde\tau\sqrt{n}} - 1\right| \leq& \ \E\left(\frac{N}{n\widetilde\tau^2} |\widehat\varsigma^2 - \widetilde\varsigma^2|\right) + \frac{1}{n\psi}\E|N - n\psi| + \frac{1}{n\psi} + \E\left(\frac{N\widehat{R}_2^2}{n\widetilde\tau^2}\right) + \E\left(\left|2\psi^{-1} + 2\widehat{R}_2 - 1\right|\frac{\sqrt{N}}{n\sqrt{\psi}\widetilde\tau}\right)\\
			& \ + (2\psi^{-1} - 1)\E\left(\frac{N|\widehat{R}_2|}{n\widetilde\tau^2}\right) + (1 - \psi)\E\left(\frac{\psi\widehat{R}_2^2 + 2|\widehat{R}_2|}{\widetilde\tau^2}\right).
		\end{align*}
		Next, using the inequality that $N\leq n$ and $\E|N - n\psi| \leq \left\{\E\left(N - n\psi\right)^2\right\}^{1/2} = \left\{n\psi(1 - \psi)\right\}^{1/2}$, we get
		\begin{align*}
			\E\left|\frac{\widehat\tau \sqrt{\widehat{n}}}{\widetilde\tau\sqrt{n}} - 1\right|\leq& \ \E\left(\frac{1}{\widetilde\tau^2} |\widehat\varsigma^2 - \widetilde\varsigma^2|\right) + \frac{\sqrt{1 - \psi}}{\sqrt{n\psi}} + \frac{1}{n\psi} + \E\left(\frac{\widehat{R}_2^2}{\widetilde\tau^2}\right) + \E\left\{\left(2\psi^{-1} - 1 + 2|\widehat{R}_2|\right)\frac{1}{\sqrt{n\psi}\widetilde\tau}\right\}\\
			& \ + (2\psi^{-1} - 1)\E\left(\frac{|\widehat{R}_2|}{\widetilde\tau^2}\right) + (1 - \psi)\E\left(\frac{\psi\widehat{R}_2^2 + 2|\widehat{R}_2|}{\widetilde\tau^2}\right)\\
			=& \ \E\left(\frac{1}{\widetilde\tau^2} |\widehat\varsigma^2 - \widetilde\varsigma^2|\right) + \frac{\sqrt{1 - \psi}}{\sqrt{n\psi}} + \frac{1}{n\psi} + \left\{\psi\left(1 - \psi\right) + 1\right\}\E\left(\frac{\widehat{R}_2^2}{\widetilde\tau^2}\right) + \E\left(\frac{2\psi^{-1} - 1}{\sqrt{n\psi}\widetilde\tau}\right)\\
			& \ + \E\left\{\frac{|\widehat{R}_2|}{\widetilde\tau^2}\left(\frac{2\widetilde\tau}{\sqrt{n\psi}} + 2\psi^{-1} + 1 - 2\psi\right)\right\}\\
			\leq& \ \E\left(\frac{1}{\widetilde\tau^2} |\widehat\varsigma^2 - \widetilde\varsigma^2|\right) + \frac{\sqrt{1 - \psi}}{\sqrt{n\psi}} + \frac{1}{n\psi} + \E\left(\frac{2\psi^{-3/2}}{\sqrt{n}\widetilde\tau}\right) + 2\E\left(\frac{\widehat{R}_2^2}{\widetilde\tau^2}\right)\\
			& \ + \E\left\{\frac{|\widehat{R}_2|}{\widetilde\tau^2}\left(\frac{2\widetilde\tau}{\sqrt{n\psi}} + 2\psi^{-1} + 1 - 2\psi\right)\right\}.
		\end{align*}
		Next, we obtain the asymptotic bound on the absolute difference in the cumulative functions.
		
		Let $\E|\widehat{R}_2| \lesssim n^{-2\beta}$. Following the proof of Theorem \ref{thm:normality} we have $\E|\widehat\varsigma^2 - \widetilde\varsigma^2| \lesssim n^{-1/2}$. Thus, if $\widetilde\tau \gtrsim 1$ with probability 1 and $\psi \geq \epsilon > 0$ then
		\begin{align*}
			\E\left|\frac{\widehat\tau \sqrt{\widehat{n}}}{\widetilde\tau\sqrt{n}} - 1\right|    \lesssim& \ n^{-1/2} + n^{-2\beta}.
		\end{align*}
		
		Further, if $\E\left(\frac{\rho}{\widetilde\tau^3}\right) < c$ for some finite constant $c$,
		$$ \left| \Pb\left( \frac{\widehat{n} - n }{\widehat\tau \sqrt{\widehat{n}}} \leq t   \right) - \Phi(t) \right| \lesssim n^{(1-4\beta)/2} + n^{-1/2}.$$
		\textbf{Coverage error}\\
		
		The $(1-\alpha)\%$ estimated CI for $n$ is $\widehat{n} \pm z_{\alpha/2}\widehat\tau\sqrt{\widehat{n}}$.
		\begin{align*}
			& \ \left| \Pb\left(\widehat{n} - z_{\alpha/2}\widehat\tau\sqrt{\widehat{n}} \leq n \leq \widehat{n} + z_{\alpha/2}\widehat\tau\sqrt{\widehat{n}}\right) - (1 - \alpha)\right|\\
			=& \ \left| \Pb\left(\frac{n - \widehat{n}}{\widehat\tau\sqrt{\widehat{n}}} \leq z_{\alpha/2}\right) - \Pb\left(\frac{n - \widehat{n}}{\widehat\tau\sqrt{\widehat{n}}} \leq -z_{\alpha/2}\right) - \Phi(z_{\alpha/2}) + \Phi(-z_{\alpha/2})\right|\\
			\leq& \ \left| \Pb\left(\frac{n - \widehat{n}}{\widehat\tau\sqrt{\widehat{n}}} \leq z_{\alpha/2}\right) - \Phi(z_{\alpha/2})\right| + \left|\Pb\left(\frac{n - \widehat{n}}{\widehat\tau\sqrt{\widehat{n}}} \leq -z_{\alpha/2}\right) - \Phi(-z_{\alpha/2})\right|\\
			\leq& \ \sqrt{\frac{2}{\pi}} \left\{ \frac{\sqrt{n} \psi\E|\widehat{R}_2|}{ \widetilde\tau} + |z_{\alpha/2}| \E\left(  \left|\frac{\widehat\tau\sqrt{\widehat{n}}}{\widetilde\tau\sqrt{n}}- 1  \right| \right) \right\} + \frac{2C}{\sqrt{n}} \E\left( \frac{\rho}{\widetilde\tau^3} \right)\\
			\lesssim& \ n^{(1-4\beta)/2} + n^{-1/2}.
		\end{align*}
	\end{proof}
	
	\subsection{Two lists vs multiple lists}\label{app:how_many_lists_to_use}
	
	The data-set under consideration has $K$ lists. The proposed method focuses on the conditional independence assumption of two lists ($Y_1 \ind Y_2\mid \bX$). The question is, when there are more than two lists, whether one should ignore the other $K-2$ lists (i.e. delete all rows that appear in neither list 1 nor list 2, but only in one or more of the remaining lists), or keep them. To answer this question, we evaluate the variance under these two cases. Below we present the variance of the estimated population size when $\psi$ is known.
	\begin{align*}
		\var(\widehat{n}) = \var\left( \frac{N}{\psi}\right) = \frac{n\psi(1 - \psi)}{\psi^2} = n\left(\frac{1}{\psi} - 1\right).
	\end{align*}
	
	\begin{enumerate}
		\item \textbf{Only two lists used}\\
		$\psi = \Pb(Y_1 \neq 0 \text{ or } Y_2 \neq 0)$.\\
		$\gamma(\bx) = \Pb((Y_1, Y_2)\neq (0,0)\mid \bX = \bx)$.\\
		\item \textbf{All lists used}\\
		$\psi = \Pb(Y_1 \neq 0 \text{ or } Y_2 \neq 0 \text{ or } \dots Y_K \neq 0)$.\\
		$\gamma(\bx) = \Pb( \bY\neq \bzero\mid \bX = \bx)$.\\
		The $\psi$ and the $\gamma(\bx)$ for this case are larger than the ones for the two list case above.
	\end{enumerate}
	
	It is easy to see that $\var(\widehat{n})$ is smaller when all $K$ lists are used since we observe more individuals.
	
	\section{Result with simulated data for varying total population size}
	
\begin{figure}
		\centering
		\includegraphics[scale = 0.62]{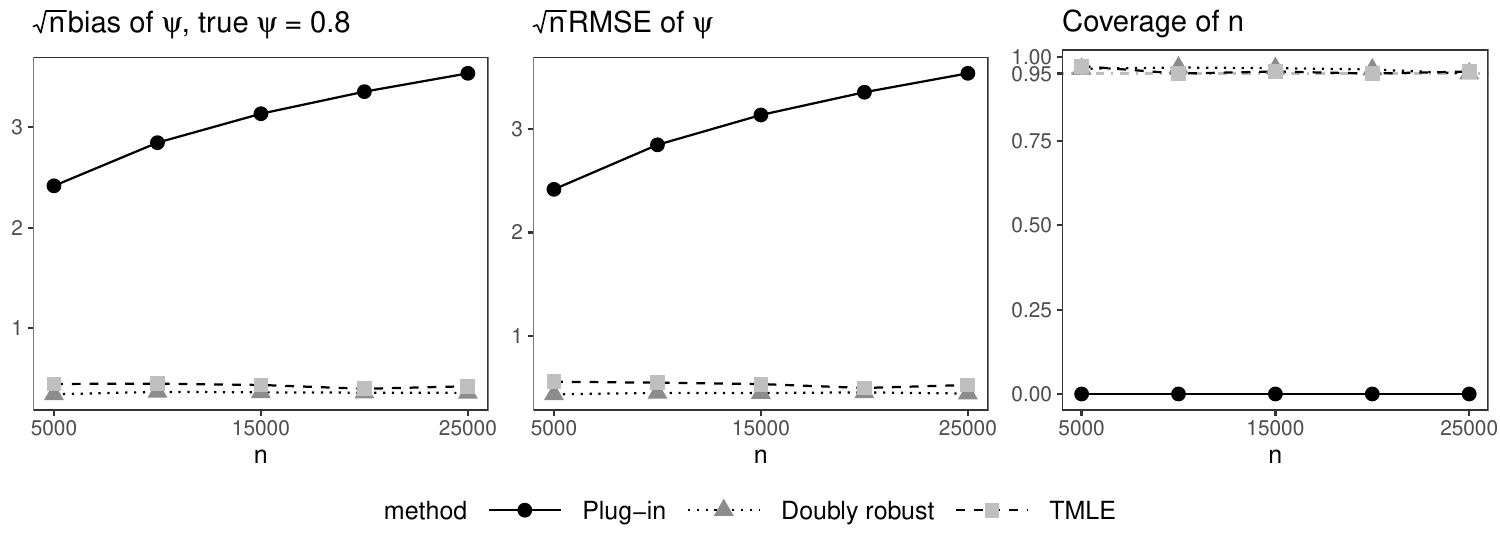}\\\vspace{-0.6cm}
		\includegraphics[scale = 0.62]{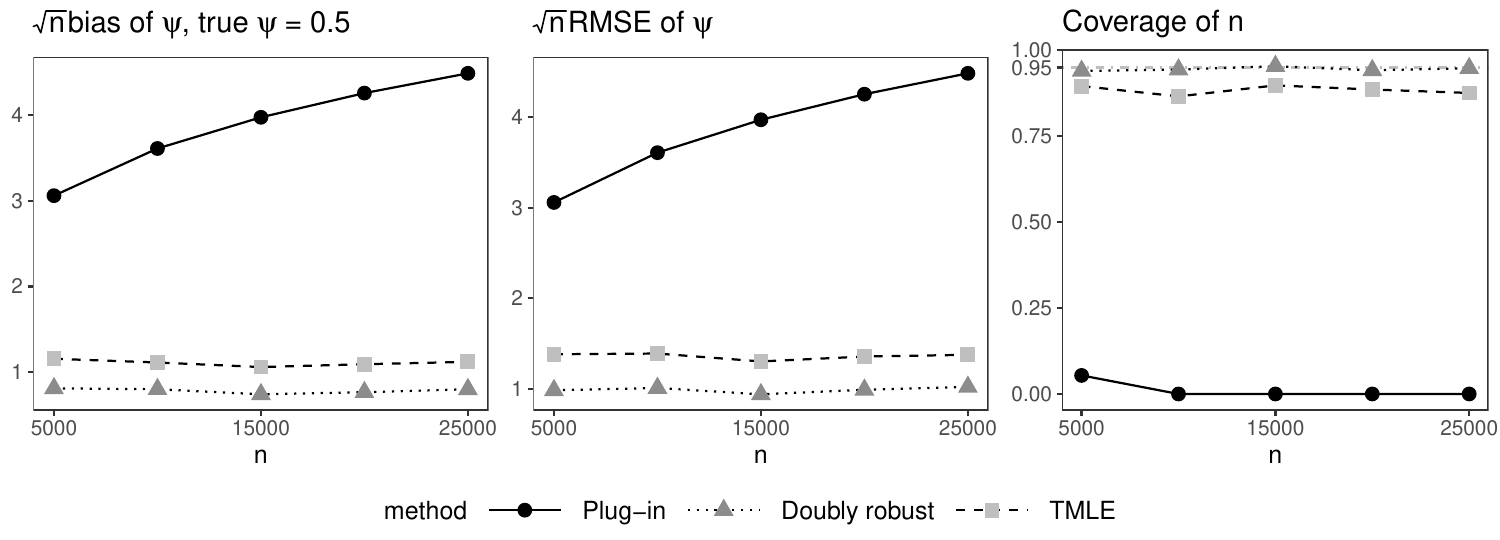}\\\vspace{-0.6cm}
		\includegraphics[scale = 0.62]{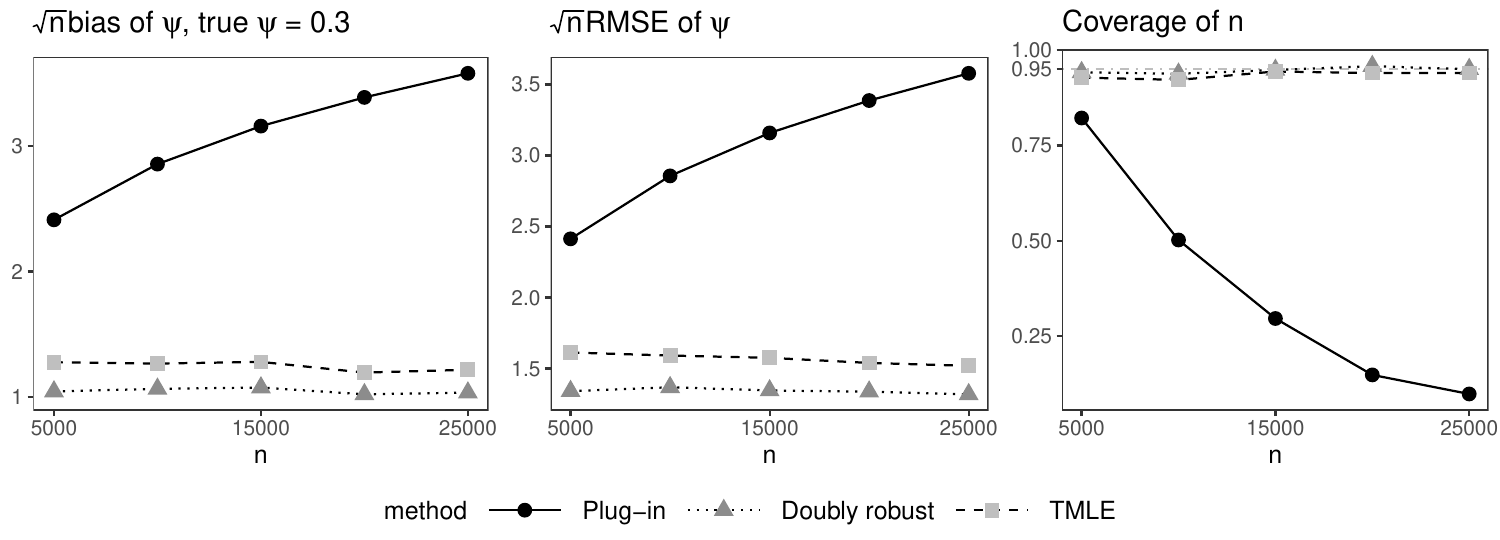}
		\caption{Estimated bias, RMSE (multiplied by $\sqrt{n}$ to study the rate), and population size empirical coverage, for simulated data with true population size $n \in \{5000, 10000, 15000, 20000, 25000\}$, across  true capture probability $\psi \in \{ 0.8 , 0.5 , 0.3\}$, and $q$-probability error rate $n^{-0.25}$.}
		\label{fig:lineplot_psi}
	\end{figure}

	The plug-in, the doubly robust and the TMLE are applied on the simulated data from section \ref{sec:simulation} total population size varying from 5000 to 25000. We focus on the case $\alpha = 0.25$ since it is the non-parametric convergence rate. The plots are presented in figure \ref{fig:lineplot_psi}. For each combination of $(\psi, n)$, we simulated a dataset 500 times. The estimated bias and RMSE times $\sqrt{n}$ shows that the doubly robust and the targeted maximum likelihood estimators have convergence rate $1/\sqrt{n}$. The plug-in estimator on the other hand, has a slower convergence rate. Similarly, the coverage of the total population size estimate for the plug-in estimator is much lower than the nominal coverage of 0.95 compared to the proposed methods' coverage.
	
	\section{Peru Internal Conflict Data 1980-2000}\label{app:realdata}
	
	The data is collected by the Truth and Reconciliation Commission of Peru \citep{ball2003many}. It was further expanded by \citet{rendon2019capturing} with the addition of geographical parameters \citep{DVN/8NXK2Y_2019}. The original dataset contains the geographic location in the form of Peru's UBIGEO codes. \citet{rendon2019capturing} added the continuous geographical coordinates for each region.
	
	\citet{ball2003many} divided Peru into 58 stratas using the UBIGEO codes. We used the approach of \citet{rendon2019capturing} and calculated the latitude, longitude and area for a region (department or strata) using shape files from \citet{peru2014geo}. For the latitude and longitude of a region, we averaged the latitude and longitude of the border of that region as in \citet{rendon2019capturing}.
	
	The victims that have no assigned strata are discarded by \citet{ball2003many}. We present the statistics of these victims in table \ref{tab:missingcount}.
	
	\begin{table}[ht]
		\centering
		\begin{tabular}{|r|rrrr|}
			\hline
			Department & State & PCP-Shining Path & Others & Unidentified \\ 
			\hline
			No department available & 396 & 137 &  11 & 369 \\ 
			Ayacucho & 1257 &  21 & 102 &  38 \\ 
			Huancavelica &  79 &   0 &   2 &   0 \\ 
			Junin &  73 &   1 &   0 &  10 \\ 
			Lima &  56 &   3 &   0 &   4 \\ 
			San Martin &  76 &   0 &   3 &   1 \\ 
			\hline
			Total & 1937 & 162 & 118 & 422 \\\hline
		\end{tabular}
		\caption{This table shows the count of the victims with missing strata information by perpetrator and department. 75\% of these victims have been captured by lists DP and ODH. 73\% of the victims belong to the State.}\label{tab:missingcount}
	\end{table}
	
	List of covariates used to model the nuisance functions are as follows:
	\begin{itemize}
		\item age: numeric variable and takes 0 for missing age.
		\item indicator for non-missing age: takes value 1 if age is present and 0 otherwise.
		\item gender: has levels male, female and others (including missing).
		\item situation: whether the victim was killed or disappeared.
		\item perpetrator: four levels indicating whether the individual is a victim of the State, Shining Path, others, or unidentified groups.
		\item indicator for non-missing department information: takes value 1 if department information is available for the individual and 0 otherwise.
		\item department latitude, longitude and area in hectares. For the individuals with missing department code, we use the average latitude, average longitude and median area of all the departments.
		\item strata code: 59 possible levels. Details and construction of the 58 strata are available in \citet{ball2003many}). Those with missing strata take value 59.
		\item strata latitude, longitude and area in hectares. For the individuals with missing strata code, we use the average latitude, average longitude and median area of all the strata.
		\item indicator of non-missing strata code.
		
	\end{itemize}
	
	\subsection{Results}
	
\begin{figure}[h]
		\centering
		\includegraphics[scale = 0.875, trim = {1cm 0 0 0}, clip]{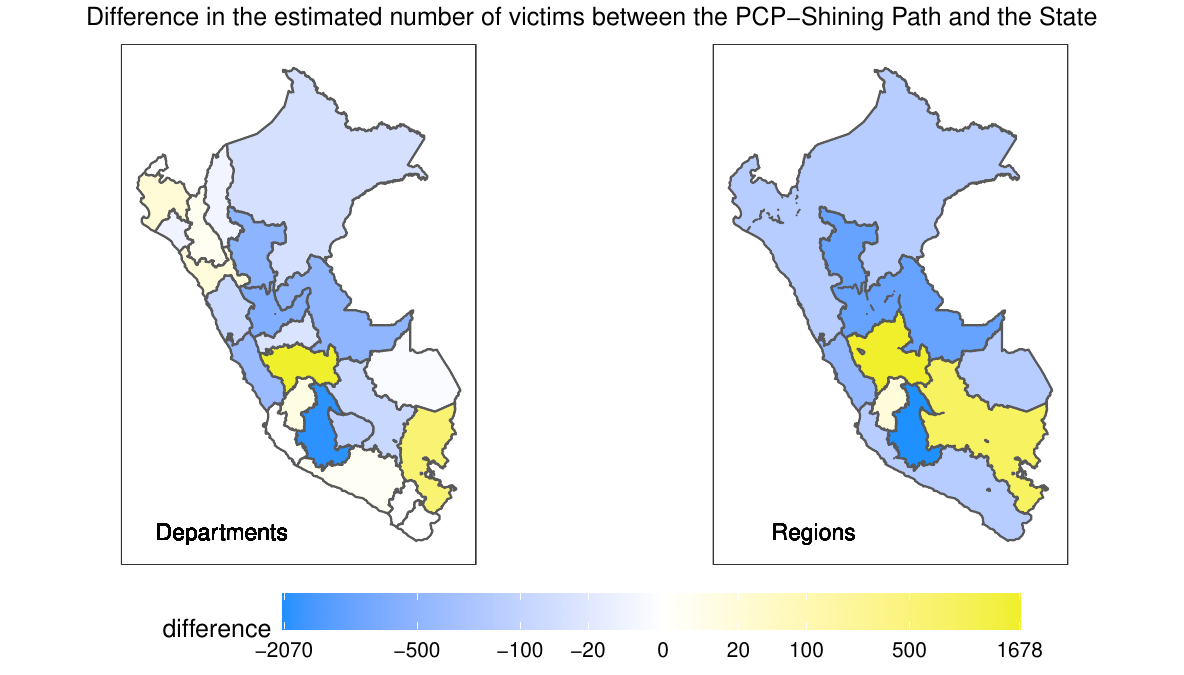}
		\caption{Difference in the estimated number of killings by the PCP-Shining Path and the State in the 25 departments and the seven regions of Peru \citep{ball2003many} from 1980-2000. The departments with comparatively higher number of victims for the State are in blue and the ones with higher number of killings for the Shining Path are in yellow.}
		\label{fig:Peru_kill_department_region}
	\end{figure}

	We present the exact estimated number of victims using our proposed doubly robust estimation in Table \ref{tab:estimatedvictims}. The difference in the estimated number of victims of the State and the Shining Path for the 25 departments and the seven geographic regions \citep[see]{ball2003many} are presented in Figure \ref{fig:Peru_kill_department_region}. The State has a significantly higher estimated number of victims in department Ayacucho and the Northern region. The Shining Path has higher estimated number of victims in departments Junin and Puno.
	
	\begin{table}[h]
		\centering
		\begin{tabular}{|lrrc|}
			\hline
			Perpetrator & N & $\widehat{n}$ & 95\% CI \\ 
			\hline
			State & 11,564 & 20,756 & [13,775, 27,737] \\ 
			PCP-Shining Path & 9,243 & 13,313 & [10,333, 16,293] \\ 
			Unidentified & 3,399 & 25,749 & [13,384, 38,114] \\ 
			Total & 24,692 & 68,874 & [58,543, 79,204] \\ 
			\hline
		\end{tabular}
		\caption{Observed and estimated numbers of killings and disappearances by perpetrator, using the proposed doubly robust method, with 95\% confidence intervals.}\label{tab:estimatedvictims}
	\end{table}
	
\end{document}